\documentclass[aps,prd,nofootinbib,twocolumn,superscriptaddress,preprintnumbers,balancelastpage,longbibliography]{revtex4-2}
\usepackage{aas_macros}

\usepackage{placeins}
\usepackage{amsmath,amssymb,mathtools,bm}
\usepackage{graphicx, color, hepunits}
\usepackage[dvipsnames]{xcolor}
\usepackage{cancel}
\usepackage[normalem]{ulem}
\usepackage{float}
\usepackage{filecontents}
\usepackage{multirow}
 \usepackage{hyperref} 
\hypersetup{
    colorlinks=true,       
    linkcolor=blue,        
    citecolor=blue,        
    filecolor=magenta,     
    urlcolor=blue          
}
\usepackage[utf8]{inputenc}
\usepackage[english]{babel}
\usepackage{array}

\usepackage{tikz}
\usepackage{tikz-feynman}
\tikzfeynmanset{compat=1.1.0}

\newcommand{\gagg}{g_{a \gamma \gamma}}
\newcommand{\gaee}{g_{aee}}

\newcommand{\gaeegagg}{g_{aee} \times g_{a \gamma \gamma}}

\begin{document}

\title{Probing the Axion-Electron Coupling with NuSTAR Observations of Galaxies}

\author{Orion Ning}
\affiliation{Berkeley Center for Theoretical Physics, University of California, Berkeley, CA 94720, U.S.A.}
\affiliation{Theoretical Physics Group, Lawrence Berkeley National Laboratory, Berkeley, CA 94720, U.S.A.}

\author{Benjamin R. Safdi}
\affiliation{Berkeley Center for Theoretical Physics, University of California, Berkeley, CA 94720, U.S.A.}
\affiliation{Theoretical Physics Group, Lawrence Berkeley National Laboratory, Berkeley, CA 94720, U.S.A.}

\date{\today}

\begin{abstract}
We search for the existence of ultralight axions coupling to electrons and photons using data from the NuSTAR telescope directed toward the galaxies M82, M87, and M31.
We focus on electron bremsstrahlung and Compton scattering for axion production in stars, summing over the stellar populations found in the target galaxies when computing the axion luminosity.
We then compute the hard X-ray signal that arises from the conversion of these axions to photons in each galaxy's magnetic fields, 
inferred from analog galaxies in  cosmological magnetohydrodynamic simulations. Analyzing NuSTAR data toward these galaxies between roughly 20 to 70 keV, we find no evidence for axions and set leading constraints on the combined axion-electron and axion-photon coupling  at the level of $|\gaeegagg| \lesssim 8.3 \times 10^{-27}$ GeV$^{-1}$ for $m_a \lesssim 10^{-10}$ eV at 95\% confidence, with M82 providing the most stringent constraints.  
\end{abstract}
\maketitle

\section{Introduction}
Ultra-light axions are pseudo-scalar particles that can couple to the Standard Model through dimension-five and higher operators and naturally emerge from string theory compactifications~\cite{Witten:1984dg,Choi:1985je,Barr:1985hk,Svrcek:2006yi,Arvanitaki:2009fg}. These hypothetical particles are closely related to the quantum chromodynamics (QCD) axion, which may solve the strong {\it CP} problem and explain the observed dark matter~\cite{Peccei:1977hh,Peccei:1977ur,Weinberg:1977ma,Wilczek:1977pj,Preskill:1982cy,Abbott:1982af,Dine:1982ah}, except that they do not couple to QCD. On the other hand, ultra-light axions can couple to quantum electrodynamics (QED) and have derivative couplings to Standard Model (SM) fermions, including electrons. In this work we set the strongest constraints to-date on the axion-electron times axion-photon coupling for ultra-light axions using X-ray observations taken with the NuSTAR Telescope. 

\begin{figure}[!t]
\centering
\includegraphics[width=0.49\textwidth]{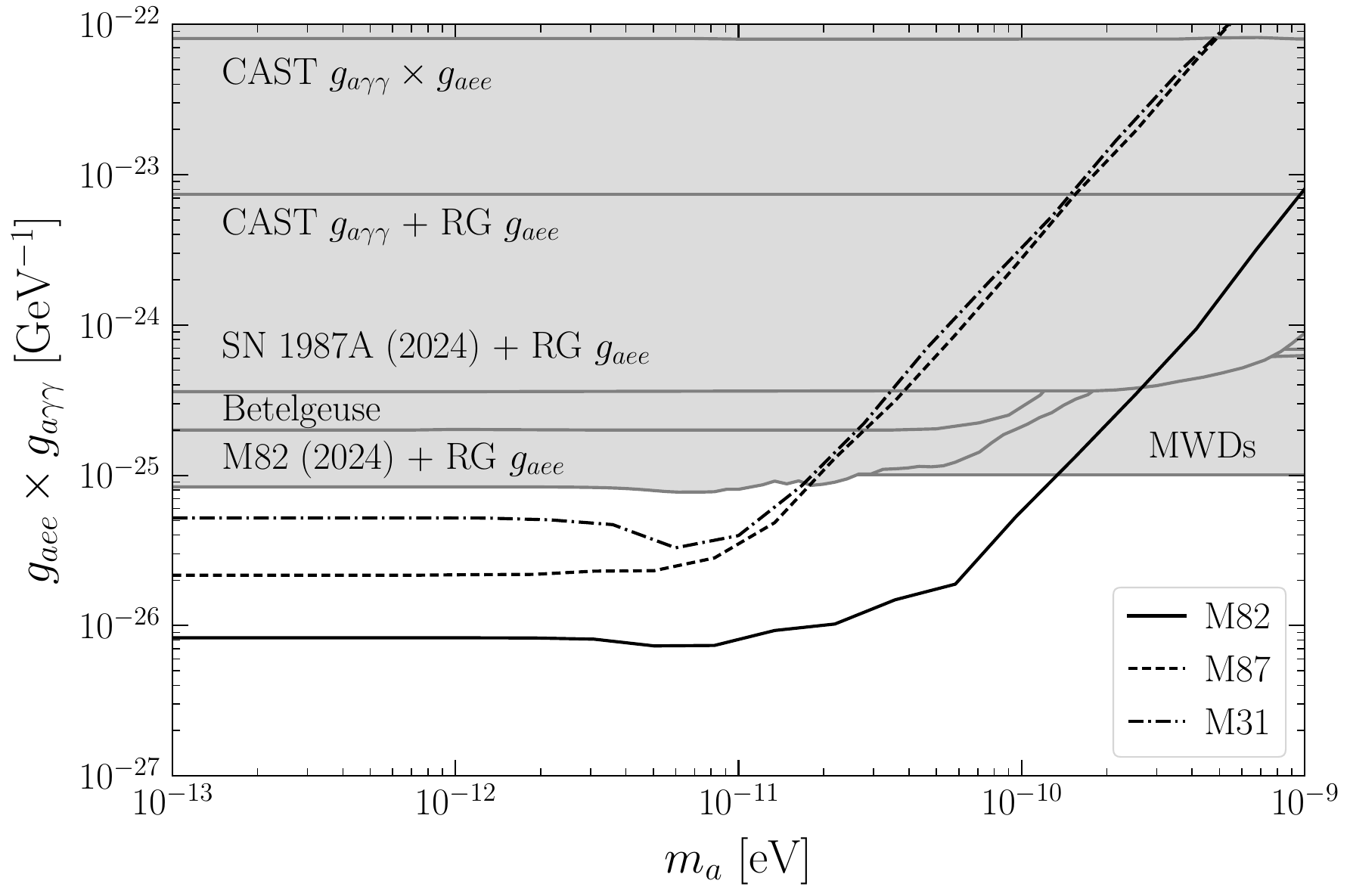}
\vspace{-0.4cm}
\caption{The parameter space of the $\gaeegagg$ coupling versus axion mass $m_a$ for ultra-light axions. The 95\% upper limits on $\gaeegagg$ from this work coming from our analyses of NuSTAR data toward M82, M87, and M31 are indicated in black. These limits are derived under the fiducial scenario in which axions are emitted from galactic stellar populations and then convert to photons in the galactic magnetic fields of each galaxy. Our upper limits at low masses surpass the direct limits previously set by CAST~\cite{2013JCAP...05..010B}, the combined limits from astrophysical probes and stellar cooling~\cite{Ning:2024eky,Manzari:2024jns,Ruz:2024gkl,  Anastassopoulos:2017ftl, Capozzi:2020cbu,Straniero:2020iyi}, as well as the limits set by NuSTAR observations of Betelgeuse (most optimistic choice in~\cite{Xiao:2022rxk}) and Chandra observations of MWDs~\cite{Ning:2024ozs, Dessert:2021bkv}.}
\label{fig:gaeegagg}
\end{figure}

The feeble axion-SM interactions can be enhanced in extreme astrophysical environments (see reviews in, \textit{e.g.},~\cite{Hook:2018dlk,DiLuzio:2020wdo,Safdi:2022xkm,Adams:2022pbo,OHare:2024nmr,Caputo:2024oqc}).  For example, axions can be produced in abundance through scattering processes in stars. Since the axions are weakly interacting, they tend to escape the stars, leading to both novel stellar cooling channels and to electromagnetic signatures if the axions convert to photons in the surrounding stellar or galactic magnetic fields.  To-date strong constraints on the axion-electron coupling have been set by modeling the effects of axion-induced cooling on the brightness of the tip of the red-giant (RG) branch~\cite{Capozzi:2020cbu,Straniero:2020iyi} and on the white dwarf (WD) luminosity function~\cite{MillerBertolami:2014rka}.  The axion-photon coupling is strongly constrained by horizontal branch (HB) star cooling~\cite{Ayala:2014pea}.  Examples of searches that look for axions produced in stars and then convert to high-energy photons in astrophysical magnetic fields include gamma-ray searches for axions produced in supernovae~\cite{Brockway:1996yr,Grifols:1996id,Payez:2014xsa,Hoof:2022xbe,Manzari:2024jns}, X-ray searches for axions produced in cooling neutron stars (NSs)~\cite{Morris:1984iz,Raffelt:1987im,Fortin:2018ehg,Fortin:2021sst,Buschmann:2019pfp}, X-ray searches for axions produced in magnetic WDs (MWDs)~\cite{Dessert:2019sgw,Dessert:2021bkv,Ning:2024ozs}, X-ray searches for axions produced in the Sun~\cite{Ruz:2024gkl}, and searches for axions produced in supermassive stars, both nearby and in the context of stellar populations~\cite{Dessert:2020lil,Xiao:2020pra,Ning:2024eky}.  These searches are also closely related to the CAST experiment ({\it e.g.},~\cite{Anastassopoulos:2017ftl}), which looks for axions produced in the Sun converting to X-rays in the magnetic field of the terrestrial experiment.  

The search we perform in this work is closely related to that of Ref.~\cite{Ning:2024eky}, which used NuSTAR hard X-ray observations of the nearby starburst galaxy M82 and of the central galaxy of the Virgo cluster, M87, to search for axions produced through the Primakoff process within stellar members of these systems converting to hard X-rays in the magnetic fields of the host galaxies and clusters.  We use the same targets and data in this work, though we also include data taken toward M31,\footnote{Another promising target we do not consider is the Galactic Center, where NuSTAR data analyses, based on conversion probabilities from~\cite{Dessert:2020lil}, would yield comparable but slightly weaker sensitivity than M31.}x but importantly here we consider axions produced through electron bremsstrahlung and Compton scattering instead of through the Primakoff process.

In this work we probe the axion coupling constant combination $g_{aee} g_{a\gamma\gamma}$, with $g_{aee}$ the axion-electron coupling and $g_{a\gamma\gamma}$ the axion-photon coupling, while Ref.~\cite{Ning:2024eky} probed $g_{a\gamma\gamma}$ alone.  The coupling constant combination $g_{aee} g_{a\gamma\gamma}$ was also probed in~\cite{Xiao:2022rxk} using NuSTAR data toward the nearby star Betelgeuse to look for axions produced in the star converting to X-rays on the magnetic fields of the Milky Way (see Fig.~\ref{fig:gaeegagg}); our work improves the sensitivity to this coupling constant combination by summing over stellar populations.  The results from this work are useful and complementary to those of~\cite{Ning:2024eky} because, depending on the Wilson coefficients of the axion effective field theory (EFT), the axion luminosity from electron bremsstrahlung and/or Compton emission could significantly outshine the luminosity from Primakoff emission.
The relevant terms in the ultra-light axion EFT for this work are 
\begin{equation}\label{eq:eft}
    \mathcal{L} \supset -\frac{1}{4} \gagg a F_{\mu \nu} \Tilde{F}^{\mu \nu} + \frac{\gaee}{2 m_e} (\partial_{\mu}a) \Bar{e} \gamma^{\mu} \gamma_5 e \,,
\end{equation}
where here $a$ refers to the axion field, $F_{\mu \nu}$ is the quantum electrodynamics (QED) field strength, $e$ is the electron field, and $m_e$ is the electron mass.  The axion-photon coupling may be written as $g_{a\gamma\gamma} = \alpha_{\rm EM} C_{a\gamma\gamma} / (2 \pi f_a)$, where $f_a$ sets the periodicity of the axion field, with {\it e.g.} $a$ periodic with period $2 \pi f_a$. The potential giving rise to the small axion mass, which is not shown but which could arise from instantons of a hidden sector or ultraviolet (UV) instantons at, for example, the scale of quantum gravity, should have this periodicity.  For theories of a single axion, $C_{a\gamma\gamma} = E$ is the integer electromagnetic anomaly coefficient, though in theories with many axions it is possible to have an additional suppression that arises from kinetic mixing~\cite{Gendler:2023kjt}.  

The axion-electron coupling may be parameterized by \mbox{$g_{aee} = C_{aee} m_e / f_a$}.  In field theory UV completions for the axion, the ratio $|C_{aee} / C_{a\gamma\gamma}|$ may vary from order unity to as small as $\sim$$10^{-4}$.  The lower limit on $|C_{aee} / C_{a\gamma\gamma}|$ arises from the fact that even if $g_{aee}$ vanishes in the UV it is generated radiatively from the coupling $g_{a\gamma\gamma}$, such that $|C_{aee}/ C_{a\gamma\gamma}| \sim \alpha_{\rm EM}^2 / \pi^2 \log(f_a / M_Z)$, with $M_Z$ the $Z$-boson mass and with the exact numerical pre-factor depending on how the axion couples independently to $SU(2)_L$ and $U(1)_Y$
(see~\cite{Dessert:2019sgw,Srednicki:1985xd,Chang:1993gm,Dessert:2021bkv}). (Note that since the $a F \tilde F$ operator is itself generated at one-loop, the radiative contribution to $C_{aee}$ is formally at two-loop order.)  On the other hand, in Dine-Fischler-Srednicki-Zhitnitsky (DFSZ)-type~\cite{Dine:1981rt,Zhitnitsky:1980tq} field theory axion models, the axion has a tree-level coupling to electrons, $|C_{aee} / C_{a\gamma\gamma}| \sim {\mathcal O}(1)$, since the electron is charged under the Peccei-Quinn (PQ) symmetry in those models.     
In string theory axion UV completions it is possible to generate one-loop-level axion-electron couplings, such that~\cite{Cicoli:2012sz,Choi:2021kuy,Choi:2024ome,Reece:2024wrn} $|C_{aee} / C_{a\gamma\gamma}| \sim (\alpha / \pi) \sim 10^{-3} - 10^{-2}$, where $\alpha$ is a  rough proxy for the fine structure constant of the SM gauge groups at high energies.

It is also possible for the ratio $|C_{aee} / C_{a\gamma\gamma}|$ to be much larger than unity. Such electrophilic axion models are not inconsistent because $C_{a\gamma\gamma}$ is not renormalized by $C_{aee}$; indeed, it is possible to have finite $C_{aee}$ and effectively vanishing $C_{a\gamma\gamma}$. Such a scenario arises, for example, in the case of ultralight axions when the SM undergoes grand unification~\cite{Agrawal:2022lsp,Agrawal:2024ejr}. When there are multiple axions in this case, the axion-photon couplings are highly suppressed for those axions with masses less than that of the QCD axion in the theory, though the axion-matter couplings need not be suppressed.  

Given that all UV completions of the axion which have non-zero $C_{a\gamma\gamma}$ also have non-zero $C_{aee}$, it is advantageous to consider probes of light axions that involve the combination of coupling constants $g_{a\gamma\gamma} \times g_{aee}$. Previous searches for this coupling combination include searches for axions produced through electron bremsstrahlung in the Sun converting to photons in the CAST experiment~\cite{2013JCAP...05..010B}, searches for axions produced again through electron bremsstrahlung in magnetic white dwarfs (MWDs) converting to X-rays in the MWD magnetospheres~\cite{Dessert:2019sgw,Dessert:2021bkv,Ning:2024ozs}, and searches for hard X-rays from Betelgeuse~\cite{Xiao:2022rxk}.  The MWD searches constrain $|g_{a\gamma\gamma} \times g_{aee}| \lesssim 1.0 \times 10^{-25}$ GeV$^{-1}$ for $m_a \ll 10^{-5}$ eV (see Fig.~\ref{fig:gaeegagg}).  In this work we improve these constraints by over an order of magnitude, for $m_a \lesssim  10^{-10}$ eV, using NuSTAR data toward M82, M87, and M31 by  accounting for axion production through electron bremsstrahlung and Compton scattering in all stars in these galaxies and then conversion to hard X-rays in the astrophysical magnetic fields permeating these systems.

\section{Axion production calculation}
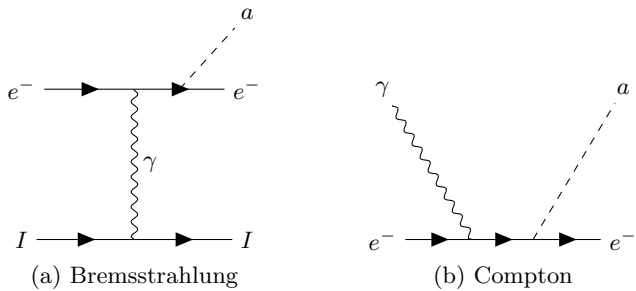
\begin{figure}[!htb]
    \centering
    \begin{tikzpicture}
        \begin{scope}[shift={(-2.5,0)}] 
            \begin{feynman}
                \vertex (i1) at (-1.5, 1) {\(e^-\)};
                \vertex (i2) at (-1.5,-1) {\(I\)};
                \vertex (a) at (0, 1);
                \vertex (b) at (0,-1);
                \vertex (c) at (0.6, 1);
                \vertex (f1) at (1.5, 1) {\(e^-\)};
                \vertex (f2) at (1.5,-1) {\(I\)};
                \vertex (f3) at (1.5, 2) {\(a\)}; 
                \diagram* {
                    (i1) -- [fermion] (a) -- [fermion] (f1),
                    (i2) -- [fermion] (b) -- [fermion] (f2),
                    (a) -- [photon, edge label=\(\gamma\)] (b),
                    (c) -- [scalar] (f3)
                };
            \end{feynman}
            \node at (0,-1.5) {(a) Bremsstrahlung}; 
        \end{scope}

        \begin{scope}[shift={(2.0,0)}] 
            \begin{feynman}
                \vertex (i1) at (-1.2, -1) {\(e^-\)};
                \vertex (i2) at (-1.2, 1) {\(\gamma\)};
                \vertex (a) at (0, -1);
                \vertex (b) at (0.8, -1);
                \vertex (f1) at (2.0, -1) {\(e^-\)};
                \vertex (f2) at (2.0, 1) {\(a\)}; 
                \diagram* {
                    (i1) -- [fermion] (a) -- [fermion] (b) -- [fermion] (f1),
                    (i2) -- [photon] (a),
                    (b) -- [scalar] (f2)
                };
            \end{feynman}
            \node at (0.4,-1.5) {(b) Compton}; 
        \end{scope}

    \end{tikzpicture}

    \caption{Feynman diagrams representing (a) the electron bremsstrahlung  and (b) Compton axion production mechanisms considered in this work, mediated by the axion-electron coupling $\gaee$.  Note that Compton emission dominates for all the galaxies considered. Additionally, here we show electron bremsstrahlung for electron-ion scattering since this dominates over electron-electron bremsstrahlung at the stellar population level.}
    \label{fig:feynmanelectron}
\end{figure}

Our focus in this work is centered on the production of axions through electron bremsstrahlung and Compton scattering occurring in M82/M87/M31's stellar populations.
The Compton axion production channel $e + \gamma \to e + a$ dominates over bremsstrahlung production at fixed $g_{aee}$, averaged over the full stellar population, even though the Compton channel requires a thermal population of photons. (Not all stars realize such a thermal photon bath.)  The bremsstrahlung channel is dominated by electrons scattering with ions $I$; {\it i.e.}, $e + I \to a + e + I$.  The axion production rates per unit volume for these processes are detailed and calculated in \cite{Raffelt:1985nk, Hoof:2021mld} and depend on the stellar temperature, density, and chemical composition.  Note that the dominance of Compton emission over bremsstrahlung in massive stars was pointed out in~\cite{Xiao:2021} in the context of axion production in Betelgeuse. 
The computations of the bremsstrahlung and Compton emissivities are presented in App.~\ref{app:bremsstrahlung}, with their Feynman diagrams illustrated in Fig.~\ref{fig:feynmanelectron}.

We extract radial profiles of $T$ and $\{n_i\}$, where $\{n_i\}$ stands for the set of ion and electron number densities, from stellar simulations using the Modules for Experiments in Stellar Astrophysics (MESA) code package~\cite{2011ApJS..192....3P,2013ApJS..208....4P}. With initial stellar mass and metallicity as the inputs, MESA evolves the star over time until its end-point, allowing us to extract the stellar profiles at the desired stellar ages. Unlike the procedure in \cite{Ning:2024eky}, we track all standard stellar populations, discussed more below, and also WDs (which we also simulate with MESA, see App.~\ref{app:stellar}). We do not track NSs, which can emit axions from electron bremsstrahlung in their crusts, because we estimate using the formalism in~\cite{Sedrakian:2018kdm} that their contribution to the axion luminosity is subdominant by at least 6 orders of magnitude relative to that from the rest of the stellar population.  

Apart from the inclusion of WDs, our stellar population modeling for M82 and M87 is otherwise identical to that presented in \cite{Ning:2024eky}. As discussed in \cite{Ning:2024eky}, M82's starburst nature implies galaxy properties that are beneficial for axion searches.  The fiducial analysis in \cite{Ning:2024eky} assumes, for M82, an initial metallicity of $Z = 0.02$, a modified Salpeter initial mass function (IMF) that flattens at low masses, a `two-burst' star formation rate (SFR) taking place across $\sim$$10^7$ years, and a total number of stars $N_{\rm tot} \sim 10^{10}$. For M87, Ref.~\cite{Ning:2024eky} adopts a similar metallicity, a canonical Salpeter IMF, and an exponential $\tau$ model with a total number of stars $N_{\rm tot} \sim 10^{12}$. Uncertainties on each of these parameters are thoroughly explored in \cite{Ning:2024eky}, with the conclusion that the uncertainties related to the assumptions used in the population modeling are ultimately subdominant to the uncertainties in the magnetic field, whose modeling will be reviewed later in this work (see also~\cite{Ning:2024eky}).  As such, in this work we adopt the fiducial population models from~\cite{Ning:2024eky} for M82 and M87. 

We also search for axions in the stellar population of M31. 
M31, situated around $\sim$$785$ kpc~\cite{2005MNRAS.356..979M} from Earth, is the nearest major galaxy to our own, and is an evolved, spiral galaxy viewed nearly edge-on~\cite{1991rc3..book.....D}. While it does not contain many young, massive stars compared to the starburst population of M82, and while it does not experience particularly strong, extended magnetic fields like M87, M31's close proximity suggests it may have competitive sensitivity to axion-induced signals, as we will see. As our fiducial model of the stellar population of M31, we adopt a Salpeter IMF, a metallicity of $Z =0.02$, and an exponential $\tau$ model with $\tau \sim$ 2 Gyr, following the prescription detailed in~\cite{2016ApJ...827....9C} and supported by, {\it e.g.},~\cite{Cook:2020}. Using the procedure in~\cite{Ning:2024eky} to infer M87's total number of stars $N_{\rm tot}$, we estimate the number of stars in M31 to be $N_{\rm tot} \sim 6.4 \times 10^{11}$ for a fiducial stellar mass of $\sim$$1.25 \times 10^{11}$ $M_{\odot}$~\cite{2012A&A...546A...4T}. We defer further details of these estimates to App.~\ref{app:stellar}, and we illustrate how uncertainties on these parameters affect our final constraints in App.~\ref{app:systematics}.  
Note that we assume that the SFR is spatially the same throughout each galaxy, although the full picture for each of our galaxies is likely to be more complex.

As in~\cite{Ning:2024eky}, we use MESA to simulate and evolve an ensemble of stars differing in initial mass, and save the stellar states in a grid of times throughout the evolution.  We then draw $N_{\rm tot}$ times from this ensemble of stellar states according to the IMF and age distributions described above. 
We classify each star with a stellar type following the procedures outlined in~\cite{Weidner:2010, Smith:2004}. These distributions also inform us of the WD population for each of our galaxies; we simulate these WDs separately and add their axion contribution to the total luminosity (see App.~\ref{app:stellar} for additional details). For M82, we find, for electron bremsstrahlung and Compton processes summed together, that the axion luminosity in the energy range of interest is dominated by red supergiant (RSG) stars, which make up $\sim$50\% of the emission (in terms of erg/s) integrated between 20 and 70 keV, and then O-type stars ($\sim$31\%) and blue supergiants (BSG) ($\sim$10\%). We note that, in the case of M82, this is roughly similar to the stars which dominate the total axion emission arising from the Primakoff effect~\cite{Ning:2024eky}.  These are mostly hot, massive stars, abundant in young starburst galaxies like M82.  
\begin{figure*}[!htb]
\centering
\includegraphics[width=\columnwidth]{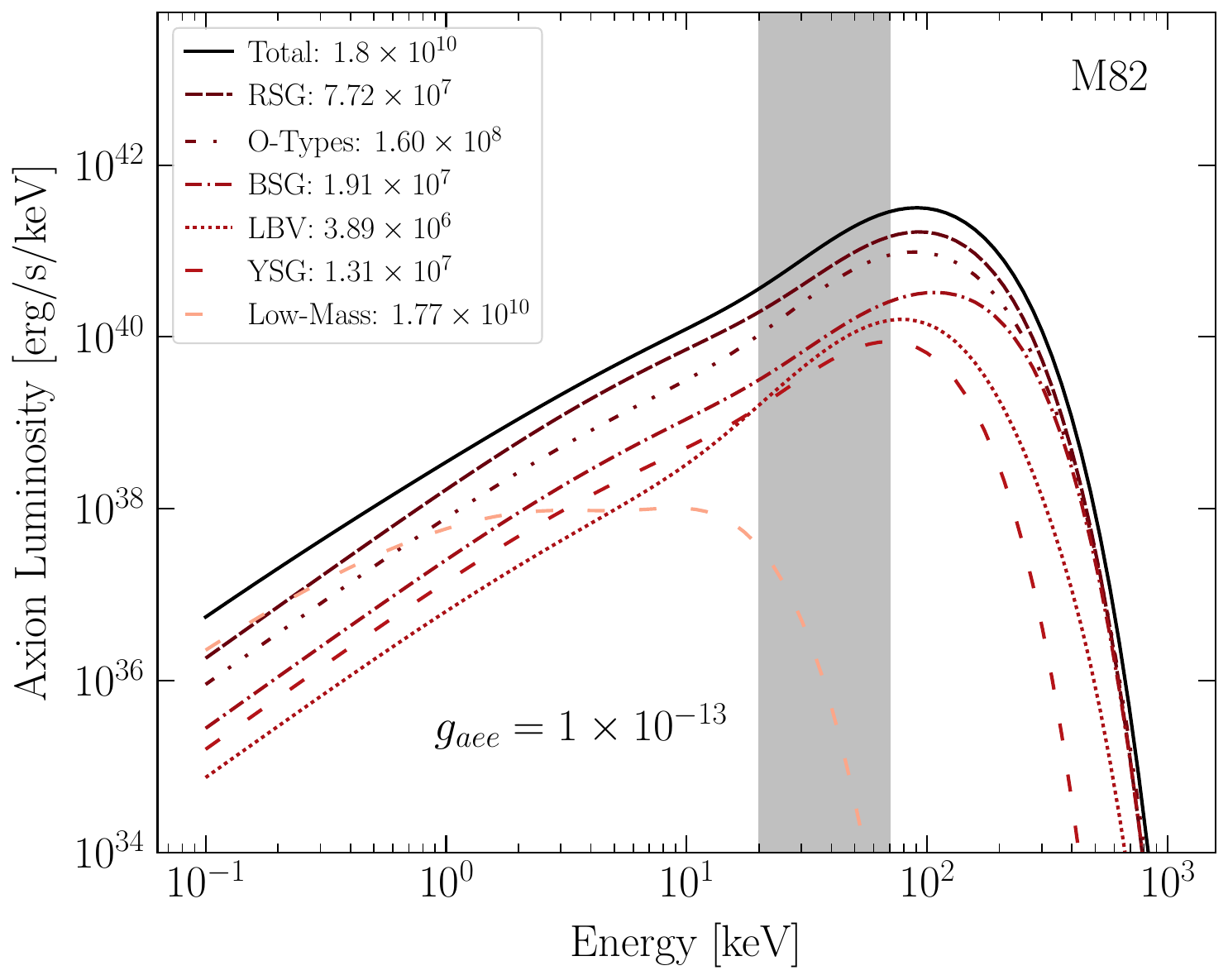}
\includegraphics[width=\columnwidth]{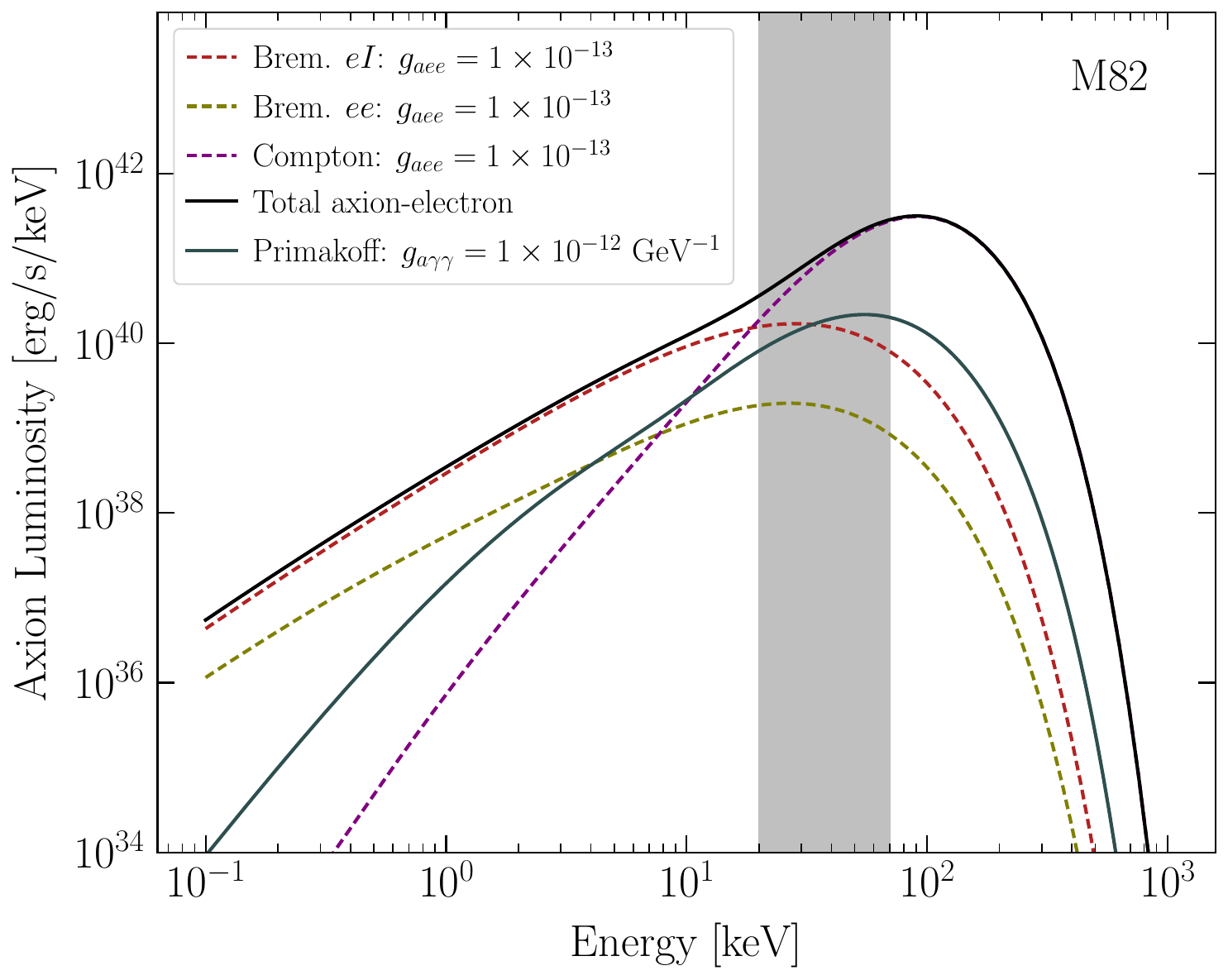}
\caption{(Left) The total axion luminosity spectrum from the  combined electron bremsstrahlung and Compton processes for the stellar population of M82 for a fixed $g_{aee} = 1 \times 10^{-13}$, with the dominant stellar classes shown separately for Red Supergiants (RSGs), Blue Supergiants (BSGs), O-type stars, Luminous Blue Variables (LBVs), Yellow Supergiants (YSGs), and all other low-mass stars.  Note that in the legend we quote the approximate expected number of stars in the galaxy for each stellar type.  (Right) A comparison of the total axion luminosity from our M82 stellar population illustrating the axion-electron processes considered in this work as well as the Primakoff effect, with the indicated couplings.}
\label{fig:M82_breakdown}
\end{figure*}
In Fig.~\ref{fig:M82_breakdown} (left), we show the differential luminosity broken down by stellar type, showing that RSGs dominate over the energy range of interest and that in fact the luminosity peaks at energies even higher than what we are able to probe with NuSTAR.  Note that for M82 WDs are subdominant compared to other stellar categories, in part because M82 does not have a large WD population given its relatively recent star formation history, though for M87 and M31 WDs play a more important (though still subdominant) role.

In the right panel of Fig.~\ref{fig:M82_breakdown} we illustrate the axion luminosity for a reference value $g_{aee} = 10^{-13}$ from the bremsstrahlung and Compton processes shown separately, indicating that Compton scattering clearly dominates the total axion luminosity, though bremsstrahlung is important at lower energies.  Moreover, the Compton emission -- and as a result the total emission from processes involving $g_{aee}$ -- is harder than the Primakoff emission, which is also shown assuming $g_{a\gamma\gamma} = 10^{-12}$ GeV$^{-1}$. (Note that an axion with these two coupling constants corresponds to $C_{aee} / C_{a\gamma\gamma} \sim 0.2$.)

For M87, our axion luminosity in the energy range of interest is dominated by a combination of RSGs ($\sim$20\%), BSGs ($\sim$30\%), and low-mass stars ($\sim$26\%) (with similar results for M31). We note that the more evolved nature of M87 and M31 imply a larger contribution from low-mass stars to the total axion emission compared to M82, as discussed further in App.~\ref{app:stellar}.

\section{ Axion-photon conversion calculation} 
After the axions are produced in the stellar interiors, we consider their conversion into photons in the galactic magnetic fields of our three galaxies. As discussed in~\cite{Ning:2024eky}, M82 has been observed to possibly host strong, turbulent, and ordered magnetic fields with field strengths on the order of $\sim$0.1 - 1 mG as a result of its starburst activity~\cite{2006ApJ...645..186T,Lacki:2013ry,2021ApJ...914...24L}, while M87 benefits from the extended magnetic fields permeating the Virgo cluster~\cite{Marsh:2017yvc}, making both galaxies ideal for our axion search. Additionally, M31 is thought to possess magnetic field strengths broadly on the order of a few $\mu$G in the outer regions to almost $\sim$20 $\mu$G near the center~\cite{Combes:1997bb, 2014A&A...571A..61G}, which also makes it a potentially strong source for axion-photon conversion for low-mass axions. 

In the presence of a magnetic field, the axion-photon interaction allows an incoming axion state to convert into a polarized electromagnetic wave (see, \textit{e.g.},~\cite{Raffelt:1987im,Safdi:2022xkm,Caputo:2024oqc}) with a conversion probability $P_{a\to \gamma}$ that depends on the axion mass $m_a$, the magnetic field,  and the plasma frequency $\omega_{pl}$ (which in turn is determined by the free-electron density $n_e$). The computation of $P_{a\to \gamma}$ requires a full-galaxy magnetic field model and $n_e$ distribution so that one can integrate along the lines of sight.  In the absence of detailed full-scale models for our galaxies considered in this work, we adopt the prescription in \cite{Ning:2024eky}, which utilized the IllustrisTNG magnetohydrodynamic cosmological simulations~\cite{Pillepich:2019bmb,Nelson:2019jkf}. These simulations allow us to obtain simulated analogue galaxies that we use to both draw stellar locations and to calculate the axion-to-photon conversion probabilities.

Here we give a broad overview of how we make use of the cosmological simulations, leaving a detailed treatment to App.~\ref{app:mag_gal} as well as the discussion in \cite{Ning:2024eky}. To extract suitable magnetic field and $n_e$ profiles for M82 and M31, we utilize the simulation data releases from the IllustrisTNG TNG50 simulations \cite{Pillepich:2019bmb,Nelson:2019jkf}, which currently comprise the highest resolution version of the IllustrisTNG simulation suite, and which have been shown to possess stellar and baryonic distributions broadly consistent with observations~\cite{Marinacci:2017wew}. Analogue galaxies are extracted from the TNG50 simulation data by identifying candidates with similar properties such as stellar mass, star formation rate (for M82), and disk morphology (for M31); the selected galaxies are then rotated to match the observed nearly edge-on orientation of M82 \cite{McKeith:1995} and M31~\cite{1991rc3..book.....D}. The procedure for M87 is similar, although we use the TNG300 simulation output in this case since we require the larger spatial scales in order to identify Virgo cluster candidates and the associated cluster-level magnetic fields, which are crucial for the axion-to-photon conversion.
 
\begin{figure}[!t]
\centering
\includegraphics[width=\columnwidth]{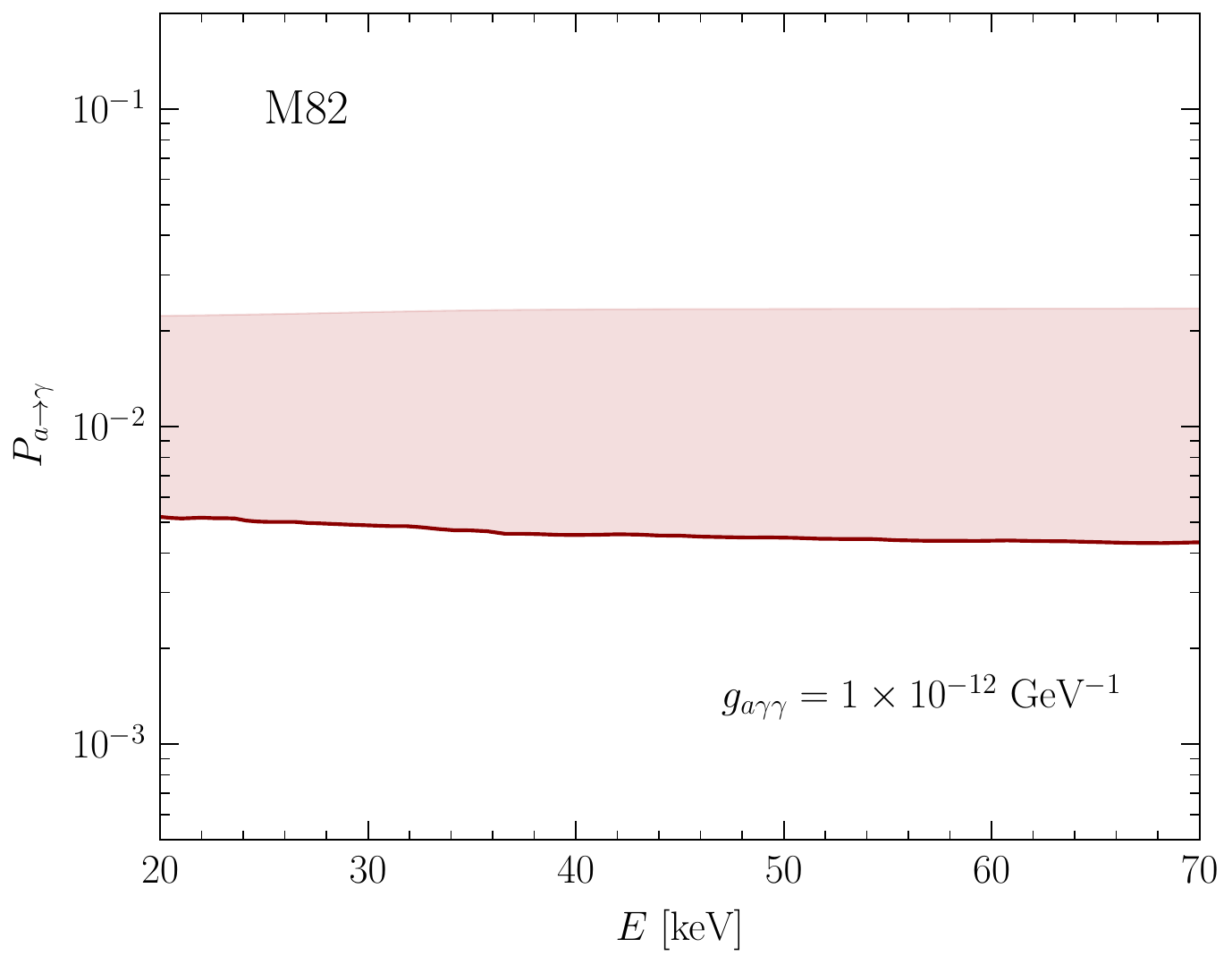}
\caption{ The fiducial average conversion probability (solid red), and the $1\sigma$ containment band illustrating the uncertainties within our analysis of the IllustrisTNG TNG50 simulation output, with the indicated $g_{a\gamma\gamma}$ and at low $m_a$.  Note that the average is computed over lines of sight originating within the stellar distributions, and the different realizations of the magnetic field model arise from taking different analogue galaxies and different galaxy orientations.}
\label{fig:M82_conv}
\end{figure}

Next, we draw stellar positions using the simulated baryonic density in our analogue galaxies as a probability distribution, and we then compute the axion-to-photon conversion probability along the line-of-sight beginning from that point. Further, since the azimuthal orientation of M82/M31 is a free parameter, we calculate the conversion probabilities for all possible azimuthal orientations of all of our galaxy candidates, and we use these probabilities to derive upper limits on $\gaee g_{a\gamma \gamma}$ for low  $m_a$. Our fiducial model is then chosen as the orientation and simulated galaxy (from our ensemble of analogue galaxy candidates) which gives the 84th percentile (\textit{i.e.}, $1\sigma$) weakest upper limit at low $m_a$. For M82, we illustrate the distribution of average probabilities $P_{a \to \gamma}$, averaged over lines of sight originating from within the distribution of baryons, in Fig.~\ref{fig:M82_conv}.
In particular, we show the 68\% containment interval, as a function of axion energy, over the ensemble of analogue galaxies and galaxy orientations. Note that our fiducial model is that for which the average conversion probability is minimized at 1$\sigma$; this model is indicated by the solid red curve. The spread in average conversion probabilities is approximately a factor of three over the ensemble of analogue galaxies and orientations.  
See App.~\ref{app:mag_gal} for further details regarding the other galaxies we consider. As discussed in \cite{Ning:2024eky} in the case of Primakoff production and as is also true here, the magnetic field uncertainty constitutes the dominant source of uncertainty in the determination of the axion signal.

\begin{figure}[!htb]
\centering
\includegraphics[width=\columnwidth]{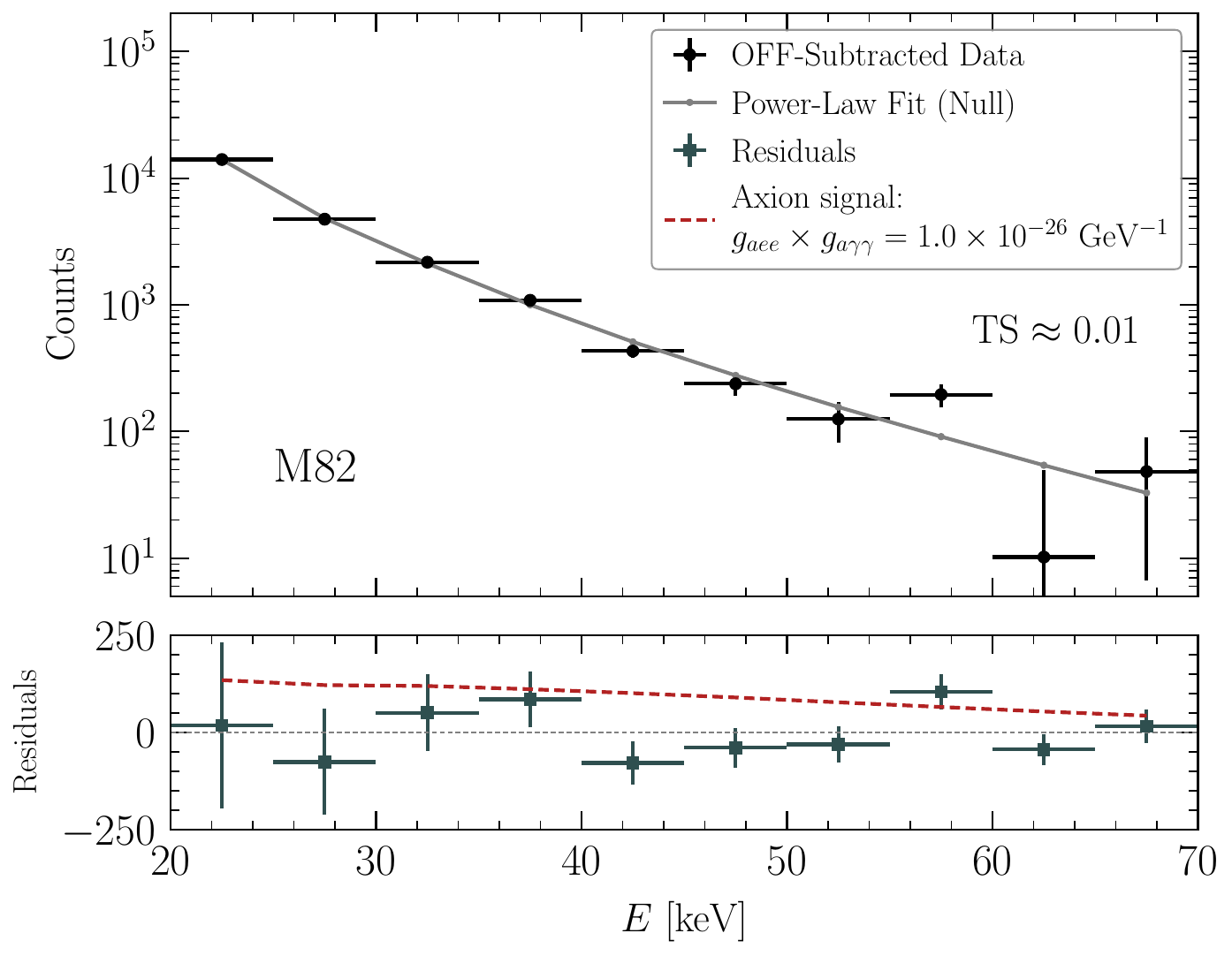}
\caption{The stacked and binned OFF-subtracted NuSTAR data toward M82 in our analysis energy range compared to our power-law null-hypothesis fit, as well as the residuals compared to an example axion-induced signal model with the indicated couplings in the low $m_a$ limit.}
\label{fig:M82_gaee_spectra}
\end{figure}

\section{NuSTAR data analysis and results}
Our NuSTAR data reduction and analysis procedures are analogous to those performed in~\cite{Ning:2024eky}. For M82 (M87), we use archival NuSTAR data amounting to $\sim$$2.07$ Ms ($\sim$$0.51$ Ms) of exposure time, for each of the two Focal Plane Modules (FPM). Using HEASoft version 6.28~\cite{2014ascl.soft08004N}, we reduce the NuSTAR data into counts spectra across source (ON) and background (OFF) regions broadly within 90\% of the NuSTAR PSF centered on our galaxy targets (for more details about these regions, see~\cite{Ning:2024eky}). We note that while Ref.~\cite{Ning:2024eky} did not analyze M31, the data reduction and analysis with NuSTAR is analogous; we use $\sim$$0.23$ Ms of archival data (for each FPM) pointed toward the central region of M31, and use similar though slightly larger ON and OFF regions than those used for M82/M87 (see App.~\ref{app:data}). After stacking data from both FPMs and rebinning in 5-keV-wide energy bins, we create a background-subtracted source spectra (the OFF-subtracted data) which we use to compare to our forward-modeled predicted axion signal.

In all cases, we compare the OFF-subtracted data to the predicted signal counts derived from forward-modeling the axion signal through the instrument response. Our axion signal model also contains a floating power-law background model (where the power-law parameters -- the normalization and spectral index -- are nuisance parameters) describing the astrophysical hard X-ray emission from our galaxies. We restrict our analysis to 20-70 keV to minimize mismodeling of low-energy X-ray emission and use a Gaussian spectral likelihood over our energy bins, profiling over our power-law nuisance parameters, in order to find the best-fit total signal model, as well as the 95\% upper limit on $|\gaeegagg|$ at fixed $m_a$ (see App.~\ref{app:data} as well as~\cite{Ning:2024eky}). An example of the OFF-subtracted NuSTAR data compared to the best-fit null model for M82 is illustrated in Fig.~\ref{fig:M82_gaee_spectra}, with more illustrations for M87 and M31 in App.~\ref{app:suppfig}.

In Fig.~\ref{fig:gaeegagg}, we show the resulting 95\% power-constrained~\cite{Cowan:2011an} upper limits on $\gaeegagg$ as a function of the axion mass  from our analyses of M82, M87, and M31. These one-sided upper limits, as well as the discovery signifances, are calculated via Wilk's theorem~\cite{Cowan:2010js}, and in App.~\ref{app:suppfig} we also illustrate the $1\sigma/2\sigma$ expectations for the 95\% upper limits under the null hypothesis, as calculated from the Asimov procedure~\cite{Cowan:2010js}. For asymptotically low $m_a$, our 95\% upper limits on $|\gaeegagg|$ for M82, M87, and M31 are $|\gaeegagg| \lesssim 8.3 \times 10^{-27}$ GeV$^{-1}$, $\lesssim 2.4 \times 10^{-26}$ GeV$^{-1}$, and $\lesssim 5.1 \times 10^{-26}$ GeV$^{-1}$, respectively. At low $m_a$, the strongest constraint on $|\gaeegagg|$ comes from the analysis of M82, with evidence in favor of the axion model being $\sim$$0.10\sigma$. In all cases, the data are consistent with the null hypothesis to within $\sim$1.5$\sigma$. Further statistics for our three galaxy analyses are illustrated in App.~\ref{app:suppfig}.

\section{Discussion}
In this work we set the strongest constraints to date on the axion-electron times axion-photon coupling for axions with masses less than approximately $10^{-10}$ eV.  We arrive at this result by computing the hard X-ray signatures coming from electron bremsstrahlung and Compton axion production in the full populations of stars in the galaxies M82, M87, and M31.  The axions undergo axion-to-photon conversion in the magnetic fields of these galaxies, and we search for the resulting X-rays in NuSTAR data. We find no evidence for axions. This work builds heavily off of~\cite{Ning:2024eky}, which performed an analogous search toward M82 and M87, but accounted for axion-production through the axion-photon coupling ({\it i.e.}, the Primakoff process).

The low-mass upper limit from this work, and in particular the analysis of M82, translates into a limit on $g_{a\gamma\gamma}$ alone of order $|g_{a\gamma\gamma}| \lesssim 1.4 \times 10^{-13} \, \, {\rm GeV}^{-1} \sqrt{\left| {C_{a\gamma\gamma} \over C_{aee}} \right|}$.  In contrast, Ref.~\cite{Ning:2024eky} set the constraint $|g_{a\gamma\gamma}| \lesssim 6.4 \times 10^{-13} \, \, {\rm GeV}^{-1}$ from the same M82 NuSTAR data.  Thus, the analyses in this work are more powerful probes of axion theories that have $|C_{aee} / C_{a\gamma\gamma}| \gtrsim 0.05$. 
 The limit from this work is dominated by the Compton process for axion production. We observe that axions emitted from the Compton process have a harder spectrum than axions emitted through Primakoff production. Future work, which may detect an axion-induced signal, could thus differentiate the two processes.

While in this work we focus on the axion-electron coupling, a similar search could be performed for the derivative axion-quark couplings, which induce axion-nucleon couplings at energies below the QCD confinement scale and which can lead to axion production in stars through nuclear bremsstrahlung and nuclear transitions. We leave a detailed study of such processes to future work. It would also be interesting in future work to consider heavy axions, which can be produced through {\it e.g.} bremsstrahlung or Compton processes in the galaxies studied here and then decay outside of the stars to photons through the axion-photon interaction (see~\cite{Benabou:2024jlj,Candon:2024eah} for similar recent works). It also remains a possibility that there are more optimal galaxy targets for axion studies in the X-ray band than the galaxies considered so far -- M82, M87, and M31 -- which may be harboring the first detectable signal of axions.

\begin{acknowledgements}
{
{\it
We thank Joshua Benabou, Chris Dessert, Joshua Foster, Maurizio Giannotti, Alessandro Mirizzi, and Yujin Park for helpful conversations. O.N. and B.R.S. are supported in part by the DOE award
DESC0025293. B.R.S. acknowledges support from the
Alfred P. Sloan Foundation.  The work of O.N. is supported in part by the NSF Graduate Research Fellowship Program under Grant DGE2146752.
This research used resources of the National Energy Research Scientific Computing Center (NERSC), a U.S. Department of Energy Office of Science User Facility located at Lawrence Berkeley National Laboratory, operated under Contract No. DE-AC02-05CH11231 using NERSC award HEP-ERCAP0023978.   
}
}
\end{acknowledgements}

\appendix

\section{Supplementary Figures}
\label{app:suppfig}

\begin{figure}[!htb]
\centering
\includegraphics[width=\columnwidth]{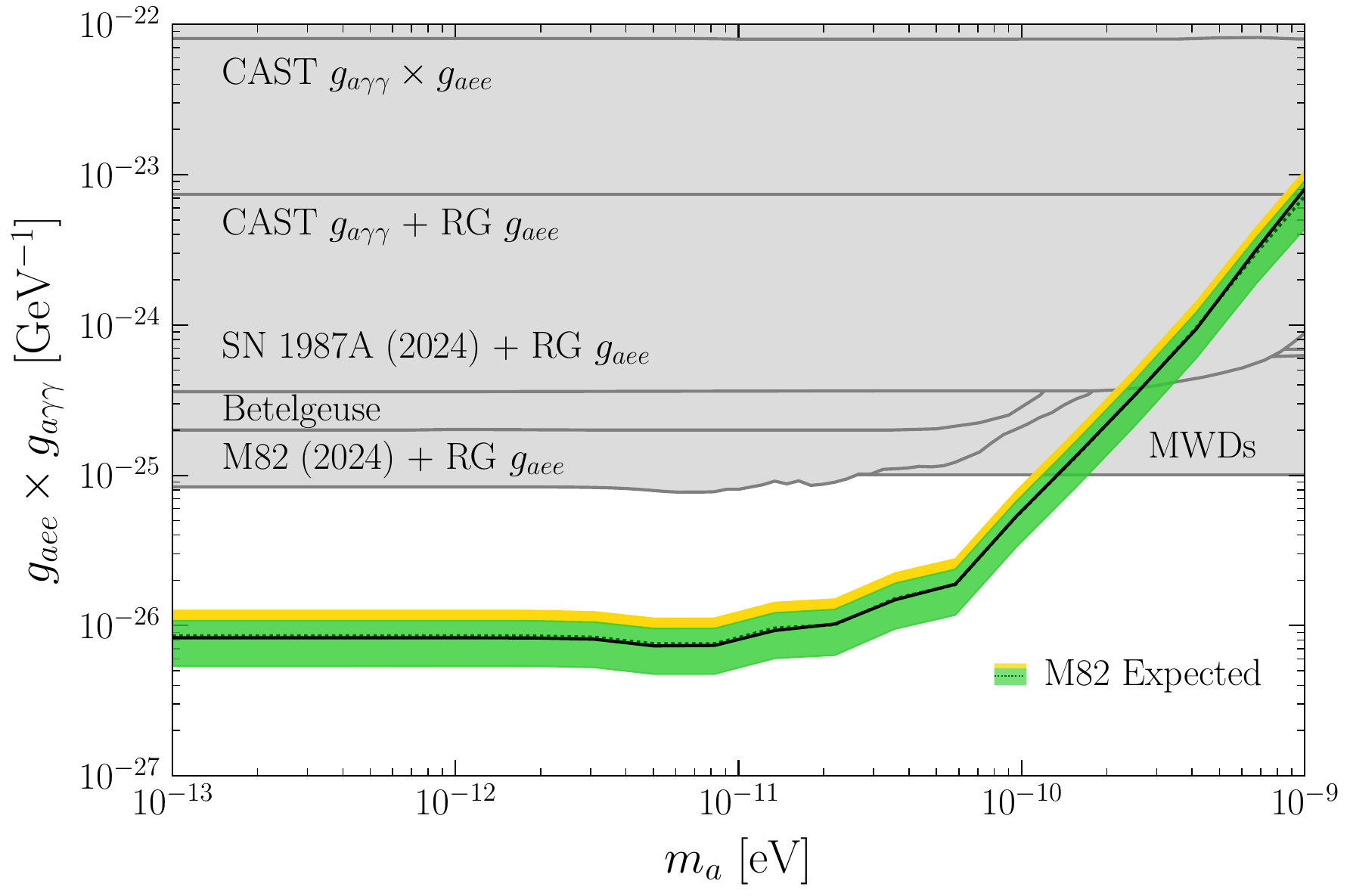}
\caption{95\% power-constrained upper limit and $1\sigma$/$2\sigma$ expectation bands under the null hypothesis for $\gaeegagg$ from the M82 analysis in this work.}\label{fig:M82_bands}
\end{figure}

\begin{figure}[!htb]
\centering
\includegraphics[width=\columnwidth]{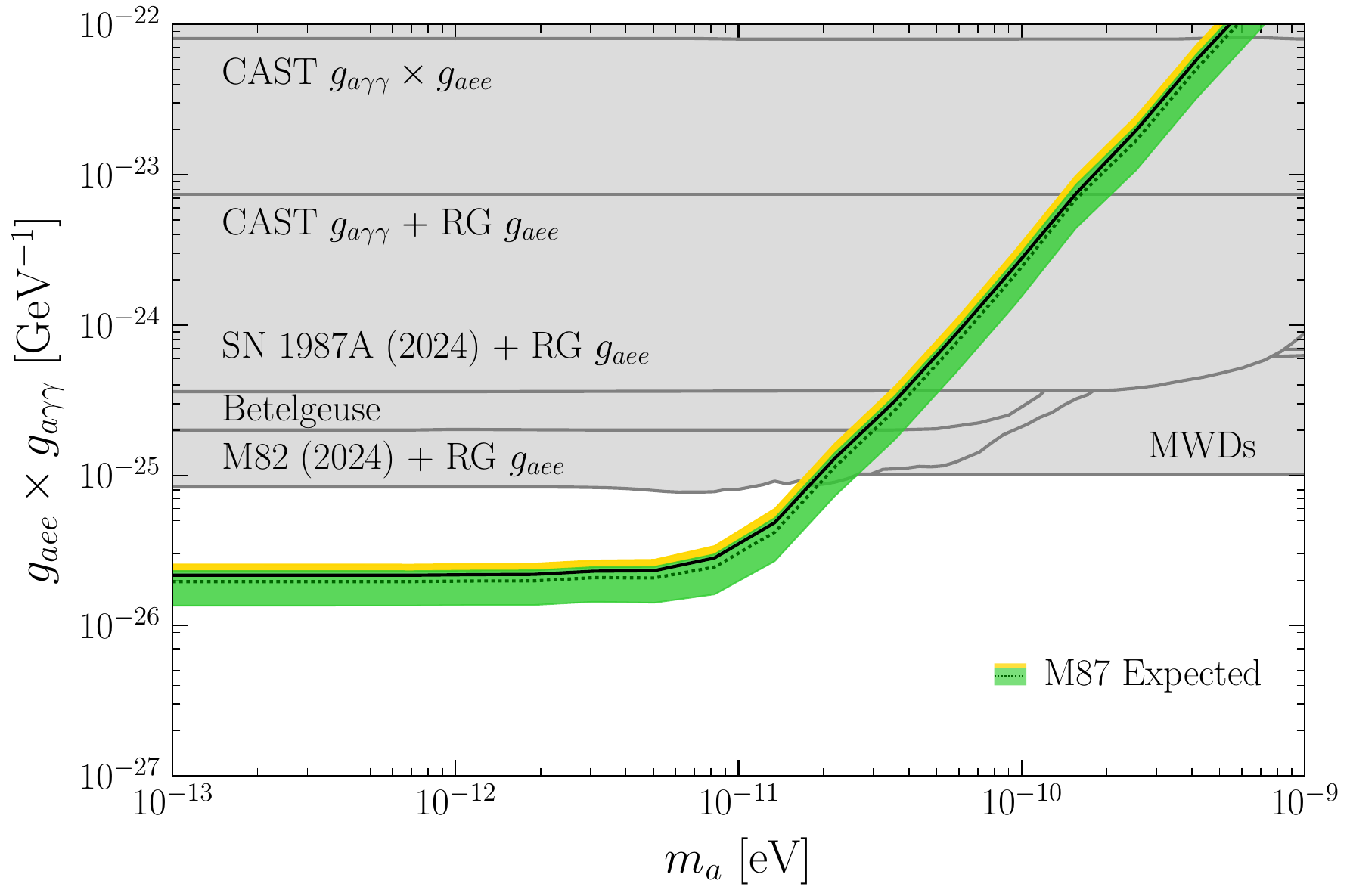}
\caption{The same as Fig.~\ref{fig:M82_bands} but for M87. }\label{fig:M87_bands}
\end{figure}

\begin{figure}[!htb]
\centering
\includegraphics[width=\columnwidth]{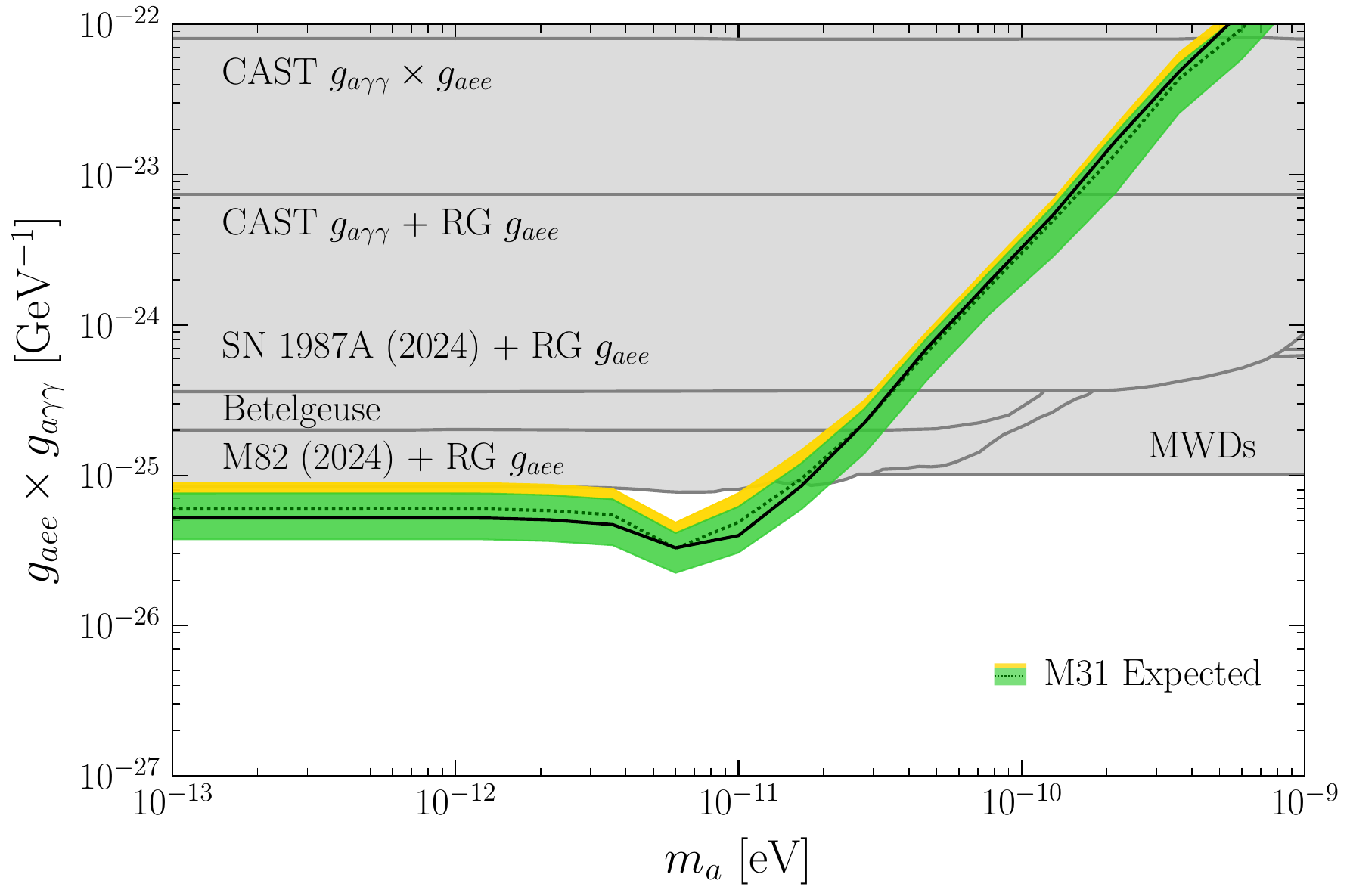}
\caption{The same as Fig.~\ref{fig:M82_bands} but for M31. }\label{fig:M31_bands}
\end{figure}

\begin{figure*}[!htb]
\centering
\includegraphics[width=\columnwidth]{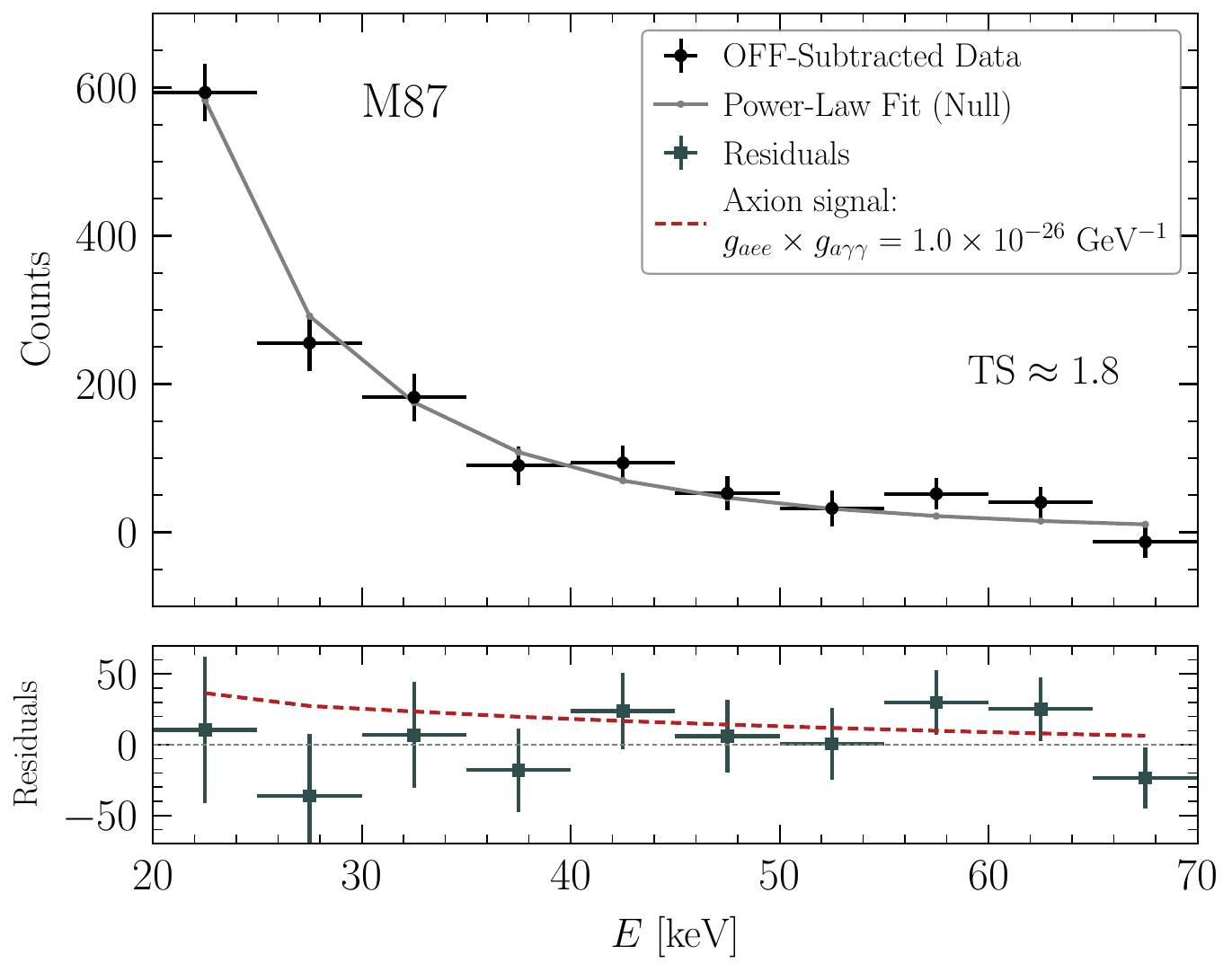}
\includegraphics[width=\columnwidth]{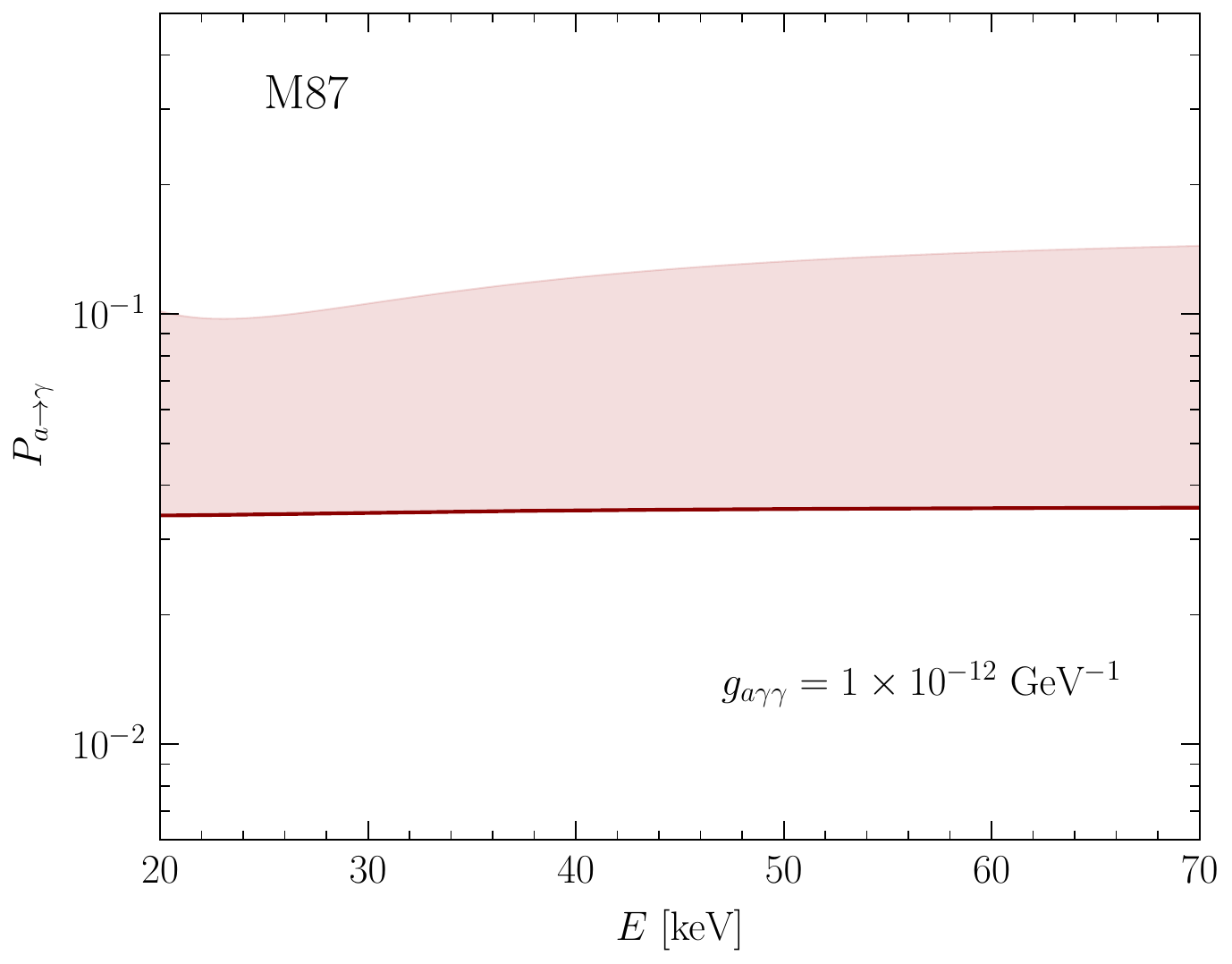}
\caption{The same as Fig.~\ref{fig:M82_gaee_spectra} (left) and Fig.~\ref{fig:M82_conv} (right) but for M87.}
\label{fig:M87_gaee_spectra}
\end{figure*}

\begin{figure*}[!htb]
\centering
\includegraphics[width=\columnwidth]{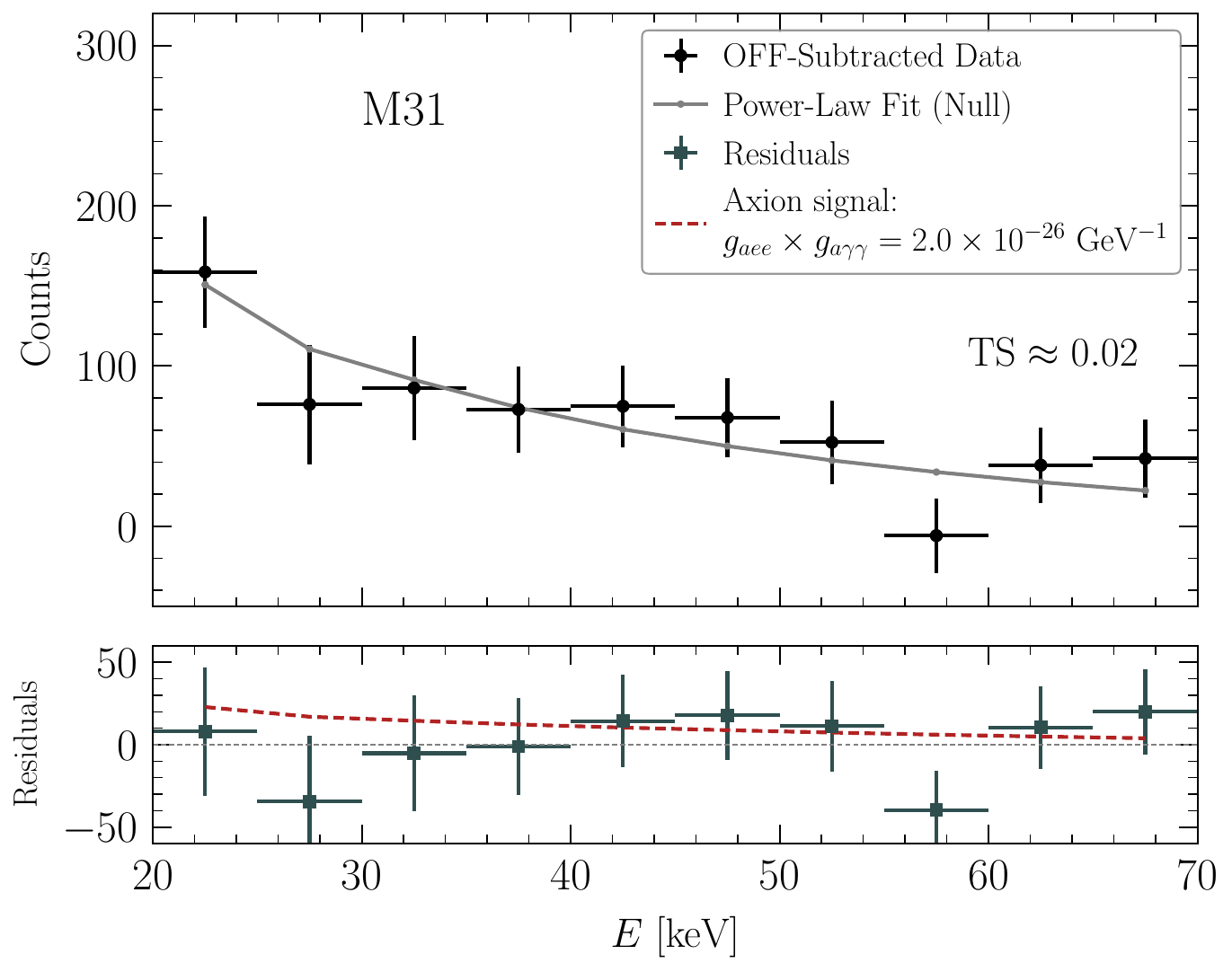}
\includegraphics[width=\columnwidth]{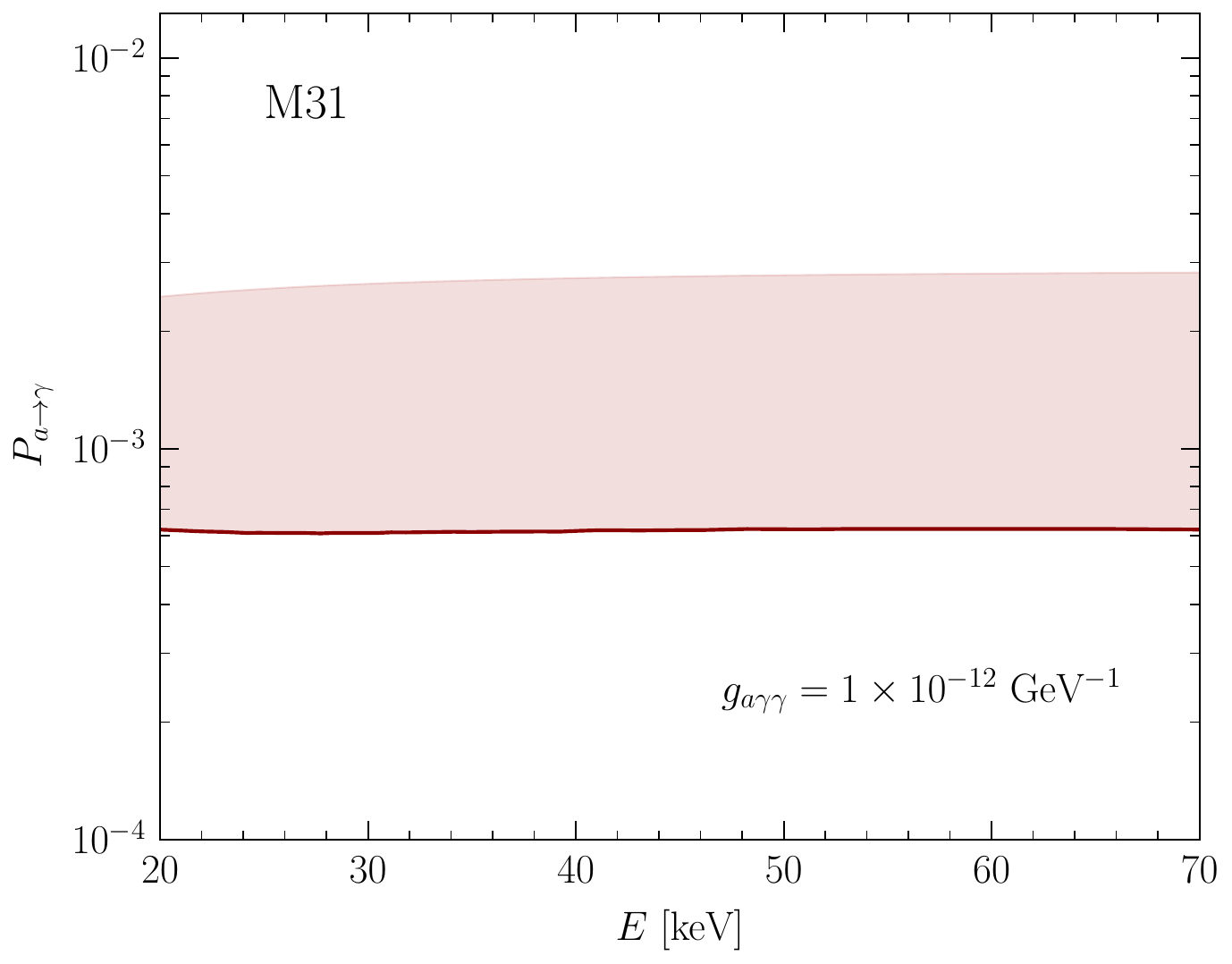}
\caption{The same as Fig.~\ref{fig:M87_gaee_spectra} but for M31.}
\label{fig:M31_gaee_spectra}
\end{figure*}

In this section we illustrate supplementary figures relevant to the results referenced in the main text. Figures~\ref{fig:M82_bands},~\ref{fig:M87_bands}, and~\ref{fig:M31_bands} illustrate the derived 95\% upper limits on $\gaeegagg$, as well as the $1\sigma$/$2\sigma$ expectations under the null hypothesis, in our analyses of M82, M87, and M31, respectively. Figures~\ref{fig:M87_gaee_spectra} and~\ref{fig:M31_gaee_spectra} compare the NuSTAR OFF-subtrated data toward M87 and M31, respectively, against the null-hypothesis fit and an example axion-induced signal model for each galaxy.

\section{Data Reduction and Analysis}
\label{app:data}

We follow an identical NuSTAR data reduction procedure as in~\cite{Ning:2024eky}, in which we use the software NuSTARDAS within HEASoft 6.28~\cite{2014ascl.soft08004N} to reprocess the data and obtain calibrated and screened events files, from which energy spectra across both focal plane modules (FPMs) and the NuSTAR field of view (FOV) are extracted. We rebin our counts spectra in 5-keV-wide bins from 5-80 keV, and also generate ancillary response files (ARFs) as well as redistribution matrix files (RMFs) which are used to forward model our axion signal. Additional details are presented in~\cite{Ning:2024eky}.

Similarly, we utilize the same archival data toward M82 and M87 as in~\cite{Ning:2024eky}. On the other hand, novel in this work is our use of NuSTAR observations toward M31. We use all archival data dedicated to the central (bulge) region of M31 to maximize the amount of stars captured in our source region, and we present the IDs and exposure times of these observations in Tab.~\ref{tab:obs}.

\begin{table}[!htb]

\begin{tabular}[t]{>{\centering\arraybackslash}p{2.5cm}>{\centering\arraybackslash}p{2.5cm}}
ObsID & $t_{\rm exp}$ [s]  \\ \hline \hline
50101001002       & $98548$       
\\
50302001006       & $42930$           
\\
50302001002       & $42355$     
\\
50302001004       & $41355$          
\\
\hline
Total & 225188
\\
\hline 
\end{tabular}
\caption{\label{tab:obs} Observation IDs and exposure time (in [s]) of the archival NuSTAR data used in our M31 analysis, in which the target is the central (bulge) region of M31.}
\end{table}

Following the prescription in~\cite{Ning:2024eky}, our axion signal is forward-modeled through the NuSTAR instrument response to return an expected number of counts coming from our putative axion signal:
\begin{equation}
    \mu^e_{S, i}(\boldsymbol{\theta}_S) = t^e \int dE' \text{RMF}_i^e(E') \text{ARF}^e(E') S(E' | \boldsymbol{\theta}_S).
\label{eq:mu_S}
\end{equation} where $t^e$ is the exposure time of the exposure $e$ in [s], and our axion signal $S$ is a spectrum of units [cts/cm$^2$/s/keV], with signal parameters $\boldsymbol{\theta}_S = \{ m_a,\gaeegagg$\}. We sum over all exposures $e$ to obtain the total number of expected signal counts across energy bins $i$.  Note that the null-hypothesis model is a power-law background model, with nuisance parameters $\boldsymbol{\theta}_B = \{ \alpha, \beta \}$ for the normalization and spectral index in flux, and this model is also folded with the instrument response.

The combined signal model plus the power-law background (with associated nuisance parameters) comprises our total predicted counts, which is then compared to our reduced NuSTAR OFF-subtracted source spectra in a Gaussian likelihood. Profiling over the nuisance parameters, we can then obtain a best-fit signal model, and through standard frequentist techniques~\cite{Cowan:2010js, Cowan:2011an} we can then obtain, given a fixed $m_a$, the 95\% upper limits on our signal parameters $\gaeegagg$ as well as the $1\sigma$ and $2\sigma$ expectations for the 95\% upper limits under the null hypothesis.

\begin{figure}[!htb]
\centering
\includegraphics[width=\columnwidth]{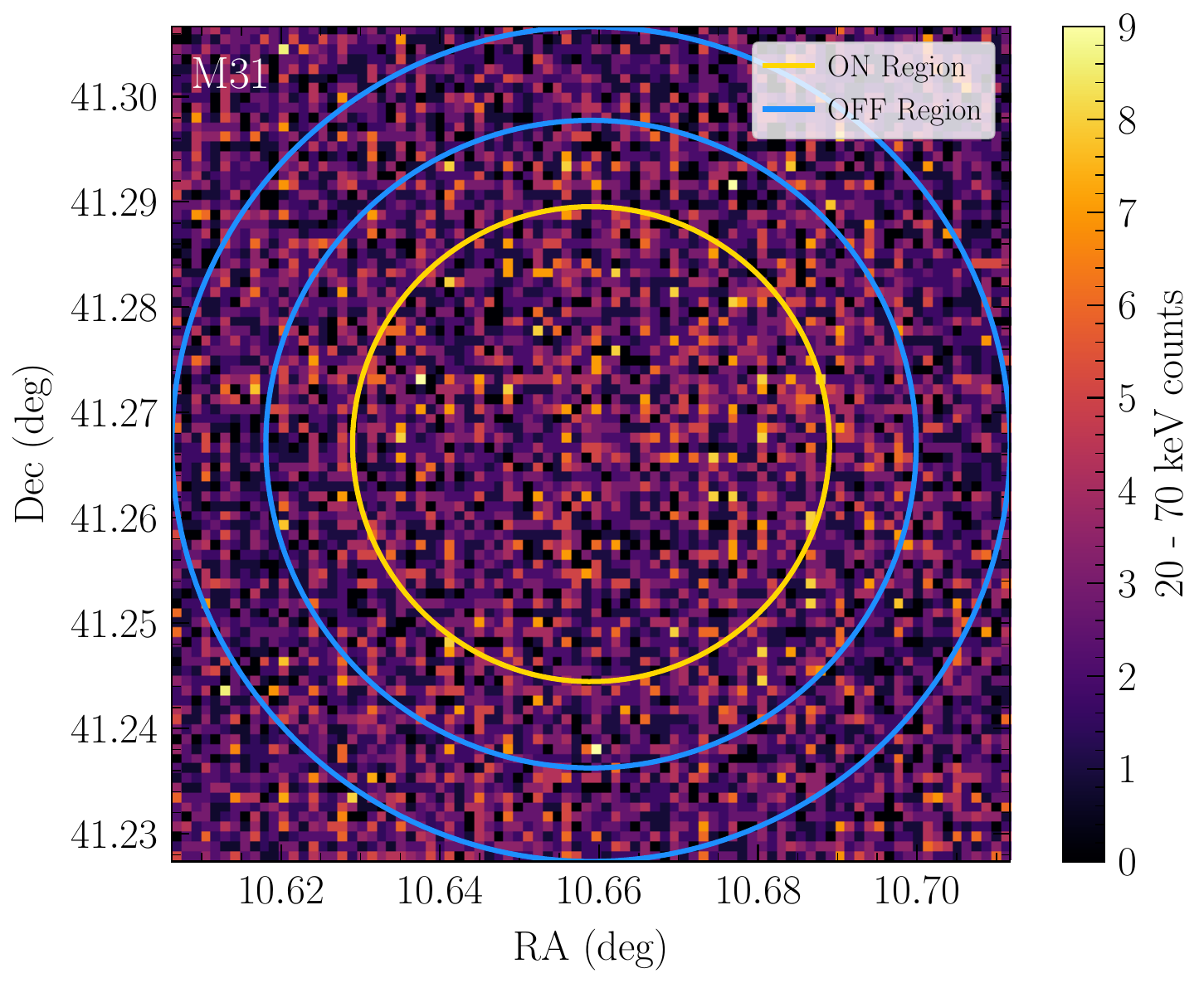}
\caption{Spatial map of the total stacked counts between 20-70 keV from all archival NuSTAR observations toward the center of M31. The hard X-ray emission seen here has a spectral morphology which is not consistent with an axion signal.}
\label{fig:M31_map}
\end{figure}

An illustration of the NuSTAR effective area for a typical observation is presented in~\cite{Ning:2024eky}, as well as the spatial maps of the total stacked counts for M82 and M87. Here, we show in Fig.~\ref{fig:M31_map} the spatial map of the data for M31, as well as the fiducial ON and OFF regions used in our analysis. As remarked in the main text for M31, our ON and OFF regions are similar though slightly larger than those used in~\cite{Ning:2024eky} to maximize the amount of stars captured by our source region. These small adjustments involve scaling our regions so that the outer background radius corresponds to the $\sim$92\% as opposed to the $\sim$90\% NuSTAR containment radius (as in~\cite{Ning:2024eky}), with the separated source region contained below the inner background radius. The NuSTAR observations involved in the map are directed toward the central (bulge) region of M31, from which one can observe hard X-ray emission whose spectral morphology we will see is inconsistent with an axion signal. Although not shown, there is clearer central emission in X-rays below the lower bound of our fiducial energy range (20 keV), which is not involved in our fiducial analysis. 

\section{Bremsstrahlung and Compton Production of Axions}
\label{app:bremsstrahlung}

In this section we provide an overview of the calculation of the axion luminosity and spectra emitted from a given star from our stellar population, under the electron bremsstrahlung and Compton production mechanisms considered in this work (also illustrated as Feynman diagrams in Fig.~\ref{fig:feynmanelectron}). In the case of electron bremsstrahlung, axions can be emitted off of electrons scattering off ions. The signal strength is controlled by the axion-electron coupling $g_{aee}$, and the differential axion luminosity per unit volume is given by~\cite{Raffelt:1985nk, Hoof:2021mld} 

\begin{equation}
{d L_a(E) \over dE} = {g_{aee}^2 \alpha_{\rm EM}^2} {\sqrt{2} \over 3 \pi^{3/2}} {(\sum_i Z_i^2 n_i) n_e \over \sqrt{T} m_e^{7/2}} {E^2 e^{-E/T}} {\mathcal{I}(E/T ,y)} \,,
\label{eq:dLadE_brem}
\end{equation}
where $Z_i$ is the charge of the scattering target nuclei, $n_i$ is the target number density, and the function $\mathcal{I}$ is defined as
\begin{equation}
{\mathcal{I}(u, y)} = \int_0^{\infty} dx x e^{-x^2} \int_{\sqrt{x^2 + u} - x}^{\sqrt{x^2 + u} + x} dt \frac{t}{(t^2 + y^2)} \,,
\end{equation}
with $y$ defined as $y \equiv \kappa / \sqrt{2 m_e T}$ and $\kappa$ the Debye screening scale (see \cite{Raffelt:1985nk}). We note that the function $\mathcal{I}$, which slightly differs from that in \cite{Raffelt:1985nk}, takes into account an updated screening description for electron bremsstrahlung processes with a modified effective form factor in the transition amplitude, as argued in \cite{Hoof:2021mld}. Additionally, axions can also be emitted off of electrons scattering off other electrons, although this form of bremsstrahlung production, which we illustrate later, is subdominant compared to scattering off of ions. The calculation is similar to that presented above, see~\cite{Hoof:2021mld}. 

Axions are predominantly  produced from Compton-like scattering, in which an incoming photon scatters with an electron to emit an axion. Again mediated by $\gaee$, the differential axion luminosity per unit volume is given by~\cite{Raffelt:1985nk, Hoof:2021mld}
\begin{equation}{
{d L_a(E) \over dE} = {g_{aee}^2 \alpha_{\rm EM}} {n_e \over 6 \pi^{2} m_e^4} {E^5 \over e^{E/T} - 1} \sqrt{1 - \frac{\omega_{\rm pl}^2}{E^2}} \,,
}\label{eq:dLadE_comp}
\end{equation}
for axion energies $E > \omega_{\rm pl}$, where $\omega_{\rm pl}$ is the photon plasma frequency.
For both~\eqref{eq:dLadE_brem} and~\eqref{eq:dLadE_comp}, the quantity $dL_a / dE$ has units of erg$/$s$/$keV$/$cm$^3$, which is then integrated over the star in order to obtain a differential luminosity.

For the case of WD stars in particular, we follow the formalism in~\cite{Dessert:2019sgw} to compute the axion emission from electron bremsstrahlung in WD interiors. Given that, in a WD core, the electrons are strongly degenerate and the temperature $T$ is much smaller than the corresponding Fermi momentum $p_F$, the emissivity in this case is essentially thermal. As in~\cite{Dessert:2019sgw} (where more details can be found), we additionally take into account screening and other medium effects through form factors $F_s$~\cite{Nakagawa:1987pga}, which are relevant for the dense plasmas affiliated with WD cores. 
 Note that there are no thermal populations of photons in WDs so there is no Compton emission of axions in these systems.

Given the above formulae for the axion luminosity of a single star, and given radial profiles for the temperatures and ion and electron number densities from MESA, we can compute the axion-induced photon spectrum at Earth. For a given star, this is calculated as
\begin{equation}
    \frac{dF}{dE}(E) = P_{a\to \gamma} (E) \frac{1}{4 \pi d^2}\frac{dL_a(E)}{dE} \,,
\label{eq:dFdE}
\end{equation}
where $d$ is the distance to the object from Earth and $P_{a\to \gamma}$ is the axion-to-photon conversion probability.

\section{Stellar Modeling and Stellar Populations}
\label{app:stellar}

Our treatment of individual and population-level stellar modeling is largely identical to that used in \cite{Ning:2024eky}. Here, we only review some of the salient aspects, and defer a detailed treatment to \cite{Ning:2024eky}.

To simulate individual stars, we use MESA (release \texttt{r23.05.1}), a one-dimensional stellar evolution code which solves the equations of stellar structure and returns detailed stellar profiles at any point in the stellar evolution. We generally use default suites and inlists for the generation of both high-mass and low-mass stars (see \cite{Ning:2024eky}), and evolve our fiducial stellar models with an initial metallicity of $Z = 0.02$, with systematics explored in App.~\ref{app:systematics}. 

MESA outputs radial profiles at many time steps along the stellar evolution, the most critical for our purposes being the temperature, density, and composition of the star. These profiles allow us to compute the axion spectrum for a given time step by integrating the axion volume emissivity ({\it i.e.},~\eqref{eq:dLadE_brem} or~\eqref{eq:dLadE_comp}) over the stellar interior. 

Our fiducial stellar population models for M82 and M87, in terms of the IMF, SFH, and number of stars $N_{\rm tot}$, are identical to those in~\cite{Ning:2024eky}. We use these models as probability distribution from which to draw MESA simulations. For each drawn simulation, we compute the axion luminosity from that star at that age using the corresponding MESA profiles, and ultimately sum all of the draws to compute the total axion luminosity emitted from the galaxy. For example, we illustrate the total axion luminosity of our M82 stellar population due to electron bremsstrahlung and Compton processes, as well as a more detailed breakdown by stellar type in Fig.~\ref{fig:M82_breakdown}, where we see that, as in~\cite{Ning:2024eky}, supergiants and O-Type stars tend to dominate.  
For M31, we adopt the modeling prescription inferred in~\cite{2016ApJ...827....9C} (as well as~\cite{Cook:2020}), which used photometric data toward M31 to constrain self-consistent stellar population modeling parameters in the pixel color-magnitude diagram framework (pCMD).
Notably, this implies an exponential $\tau$ model broadly consistent with $\tau \sim 2$ Gyr and a Salpeter IMF, although we explore reasonable variations around these values in App.~\ref{app:systematics}.

This population modeling also informs us of the age distribution of WDs available in each galaxy. As WDs are highly long-lived, old stellar remnants, we expect their contribution to be most relevant for M87 and M31, which are old, evolved galaxies with SFHs at least $\sim$$10$ Gyr. On the other hand, M82 is a very young starburst galaxy whose population is characterized by relatively large numbers of young, massive stars; it is unlikely to host substantial WD populations, given the most well-studied estimates of its SFH~\cite{ForsterSchreiber:2003ft}. Thus, in this work, we only consider the WD axion signal contributions for the populations within M87 and M31. 

Given the age distribution of WDs in our galaxies, we can draw WDs from our simulations and add their combined axion luminosity spectra to the total axion spectrum expected from a galaxy. Following the extensive WD surveys in~\cite{Kepler:2006ns} which concluded that the WD distribution is sharply peaked at $\sim$$0.6$ $M_{\odot}$, we make the simplifying assumption that our WDs are roughly $~\sim$$0.6$ $M_{\odot}$, and use a MESA model of such a WD as the template from which we draw stellar profiles based on age. We leave more complicated studies, which could incorporate more complex modeling of the poorly-understood initial-final mass relation of WDs, for future work.

\begin{figure*}[!htb]
\centering
\includegraphics[width=\columnwidth]{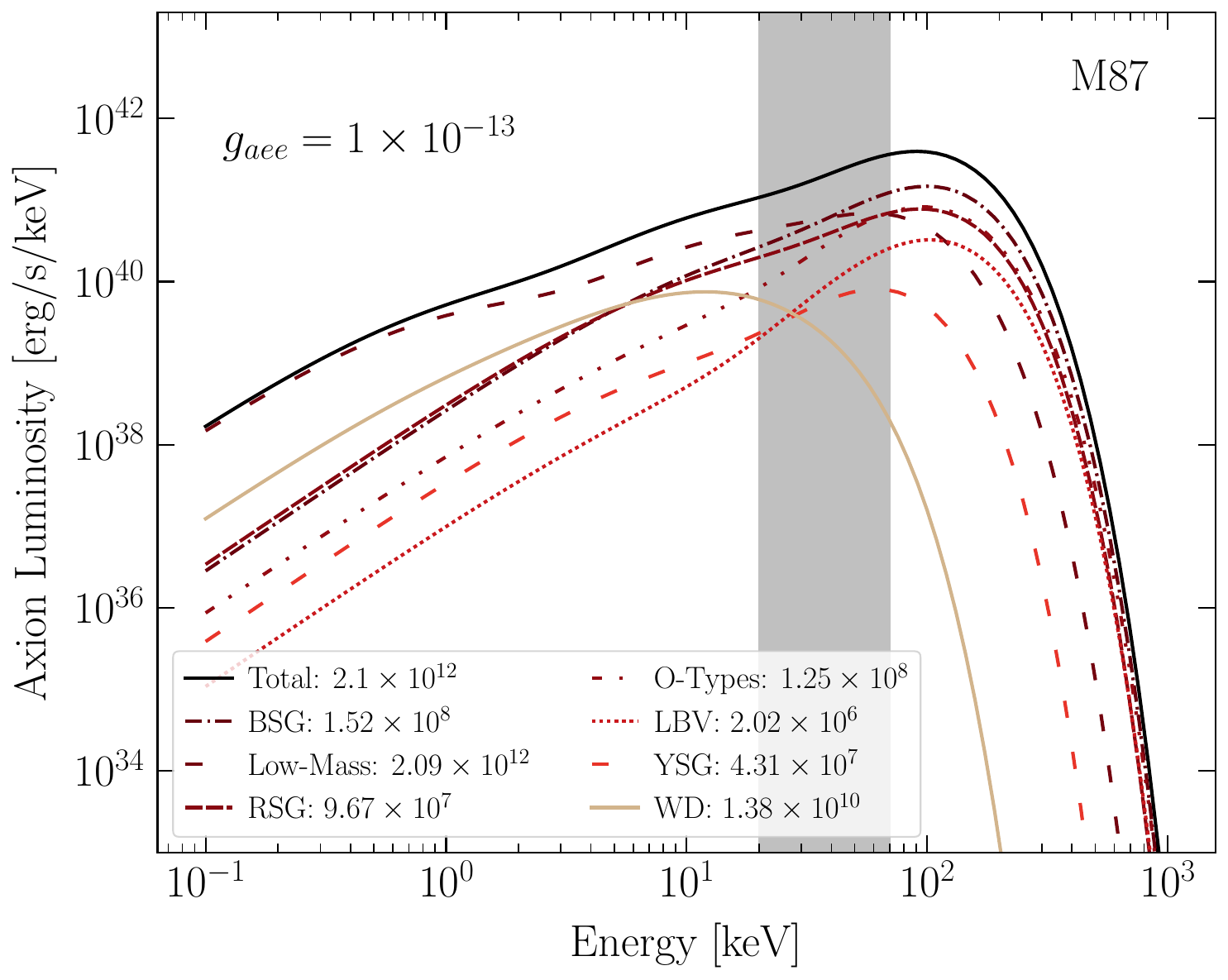}
\includegraphics[width=\columnwidth]{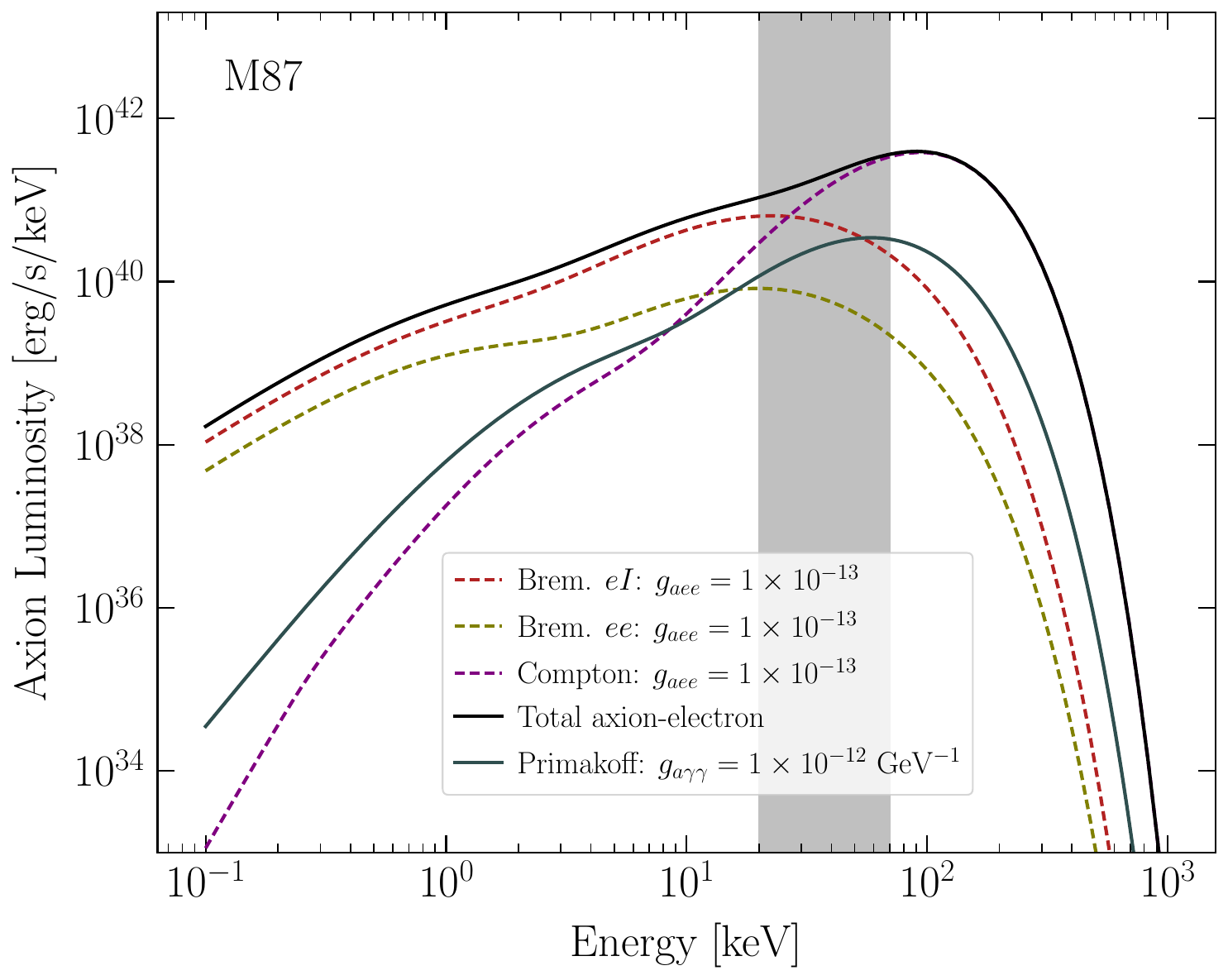}
\caption{The same as Fig.~\ref{fig:M82_breakdown} but for M87. We note the increased contribution from low-mass stars to the overall luminosity, as well as the subdominant addition of WDs.}
\label{fig:M87_breakdown}
\end{figure*}

\begin{figure*}[!htb]
\centering
\includegraphics[width=\columnwidth]{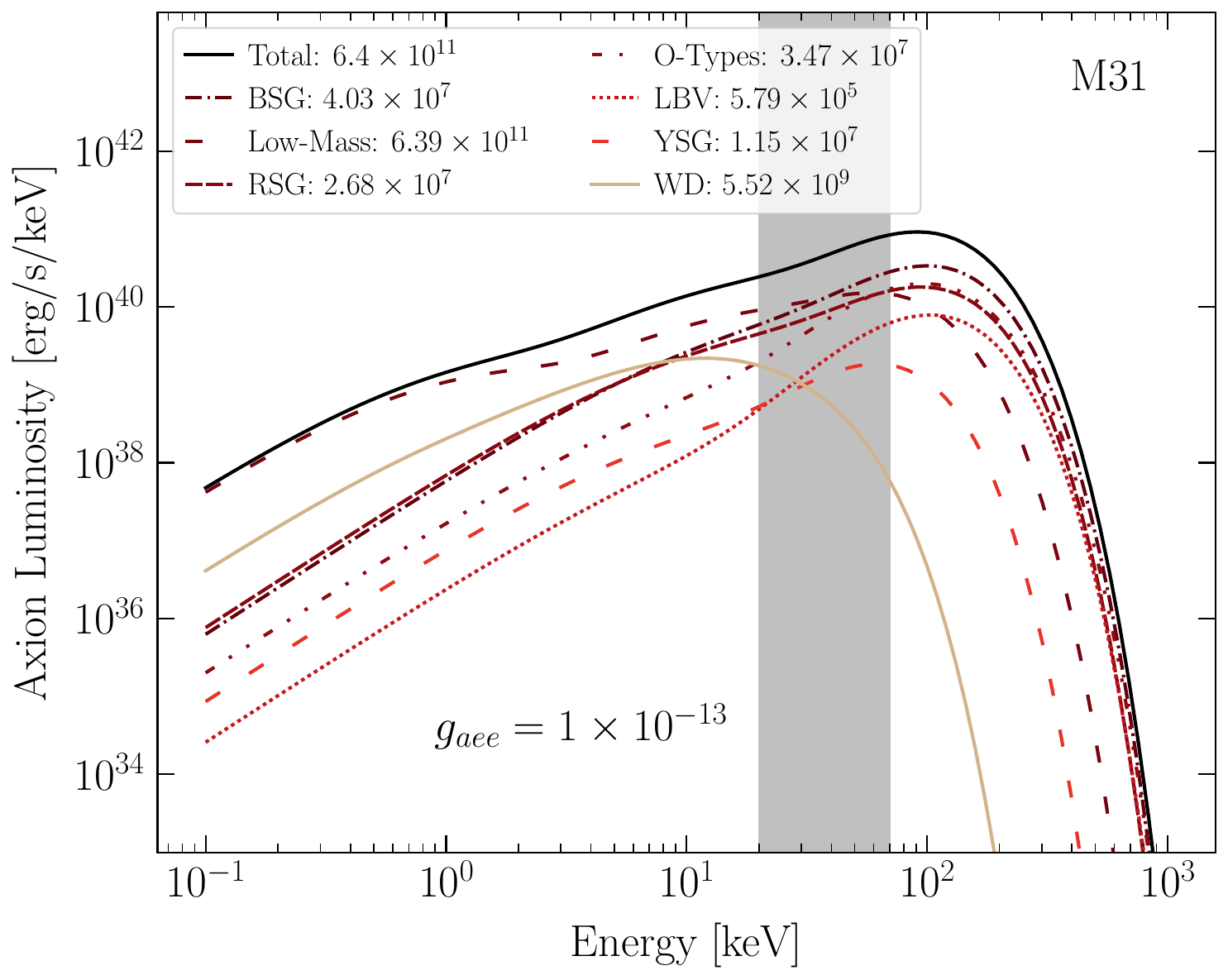}
\includegraphics[width=\columnwidth]{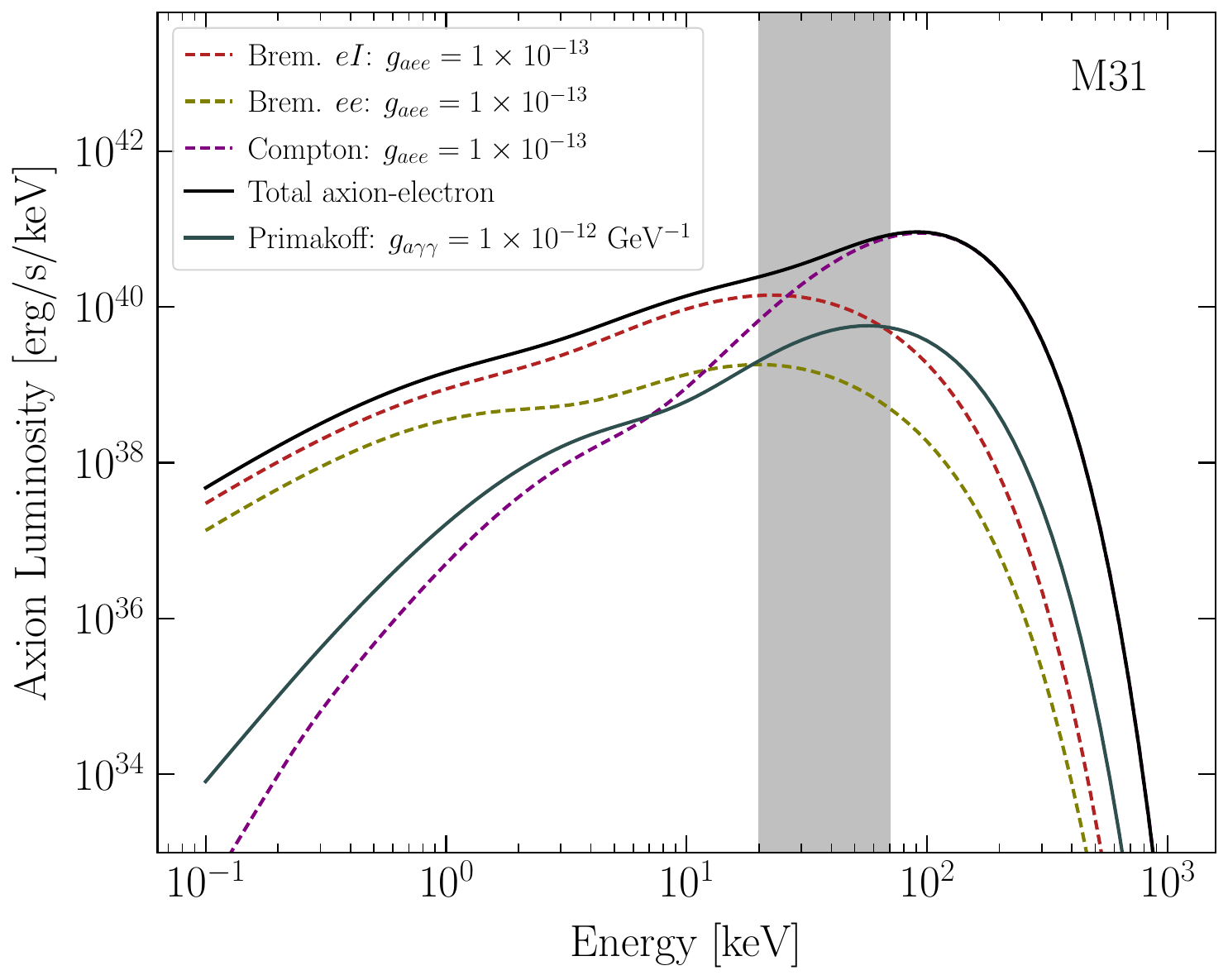}
\caption{As in Fig.~\ref{fig:M87_breakdown} but for M31. 
}
\label{fig:M31_breakdown}
\end{figure*}

We show the ultimate total axion luminosity for M87 and M31 in Figs.~\ref{fig:M87_breakdown} and~\ref{fig:M31_breakdown}. Given the similarities in the fiducial form of the SFH and IMF~\cite{Cook:2020, 2016ApJ...827....9C}, the total spectra between the two galaxies have similar morphologies. Here we see that in both galaxies, which are relatively massive and evolved, there are sizable contributions from both supergiant-like stars, as well as from low-mass stars, which tend to dominate at lower energies.  The contributions from WDs are subdominant.

\section{Magnetic Field Model and Conversion Probabilities}
\label{app:mag_gal}

We consider the conversion of axions to photons in the galactic magnetic fields of our three galaxies, following the prescription in~\cite{Ning:2024eky}. In the presence of a magnetic field ${\bf B}(s)$ and a free-electron density $n_e(s)$, an axion of mass $m_a$ and energy $E$ has a probability of converting to a photon over some line-of-sight distance $d$ which may be computed as
\begin{equation}
P_{a \to \gamma} \approx {g_{a\gamma\gamma}^2 \over 4} \sum_{j=1,2} \left| \int_0^d ds' B_j(s') e^{i \Delta_a s' - i \int_0^{s'} ds'' \Delta_{||}(s'')}\right|^2 \,, 
\label{eq:P_a_to_gamma}
\end{equation}
in the limit $P_{a \to \gamma} \ll 1$~\cite{Raffelt:1987im, Dessert:2020lil, Safdi:2022xkm}. Here, $\Delta_a \equiv -m_a^2/(2E)$, $\Delta_{||}(s) \equiv -\omega_{\rm pl}^2/(2E)$ where $\omega_{\rm pl}$ depends on $n_e$, $j$ runs over the two transverse directions, and the line-of-sight distance $s$ runs from 0 to $d$. Non-zero $m_a$ and  plasma frequency $\omega_{\rm pl}$ contribute to the phase difference picked up between the axion wave and the electromagnetic wave over the distance $d$. We use~\eqref{eq:P_a_to_gamma}, together with~\eqref{eq:dLadE_brem} (or~\eqref{eq:dLadE_comp}), to ultimately derive the expected axion-induced flux at Earth~\eqref{eq:dFdE}. Additionally, as in~\cite{Ning:2024eky}, we use the code package \texttt{gammaALPS}~\cite{Meyer:2021} to calculate the conversion probabilities~\eqref{eq:P_a_to_gamma}. 

As in~\cite{Ning:2024eky}, we use full-scale galaxy/cluster models of M82/M87 from the IllustrisTNG TNG50 and TNG300 simulation output~\cite{Pillepich:2019bmb, Nelson:2019jkf} to infer magnetic field and free-electron density models to use in~\eqref{eq:P_a_to_gamma}.
The identification of suitable candidate galaxies for M82 (clusters for M87/Virgo) was discussed in~\cite{Ning:2024eky}, and mainly involves selecting for galaxies/clusters which fall within the expected stellar mass of our targets. Then, for each candidate galaxy/cluster, we calculate upper limits over a variety of orientations (as mentioned in the main text), ultimately choosing as our fiducial model the candidate and orientation which produces the weakest upper limit at the $1\sigma$ level. We extend this methodology to M31. In the case of M31, we select for candidates whose stellar masses fall between $1.0\times 10^{11}$ and $1.5 \times 10^{11}$ $M_{\odot}$~\cite{2012A&A...546A...4T}, and, following the prescription for selecting M31 analogues broadly suggested in~\cite{2024MNRAS.535.1721P}, also filter for the correct morphology of our candidates by searching for those with disk scale lengths between $\sim$$4.8$ and $\sim$$6.8$ kpc. Ultimately, we find 6 suitable candidates with subhalo indices 388544, 414917, 432106, 448830, 458470, 461785 in the TNG50 database, which we utilize in this work.

For all three of our galaxies, uncertainties on the magnetic field modeling from IllustrisTNG are translated to uncertainties on the 95\% upper limit $\gaeegagg$, illustrated in App.~\ref{app:systematics}. Just as in~\cite{Ning:2024eky}, these magnetic field uncertainties constitute the dominant source of uncertainty in our analyses.

\begin{figure}[!htb]
\centering
\includegraphics[width=\columnwidth]{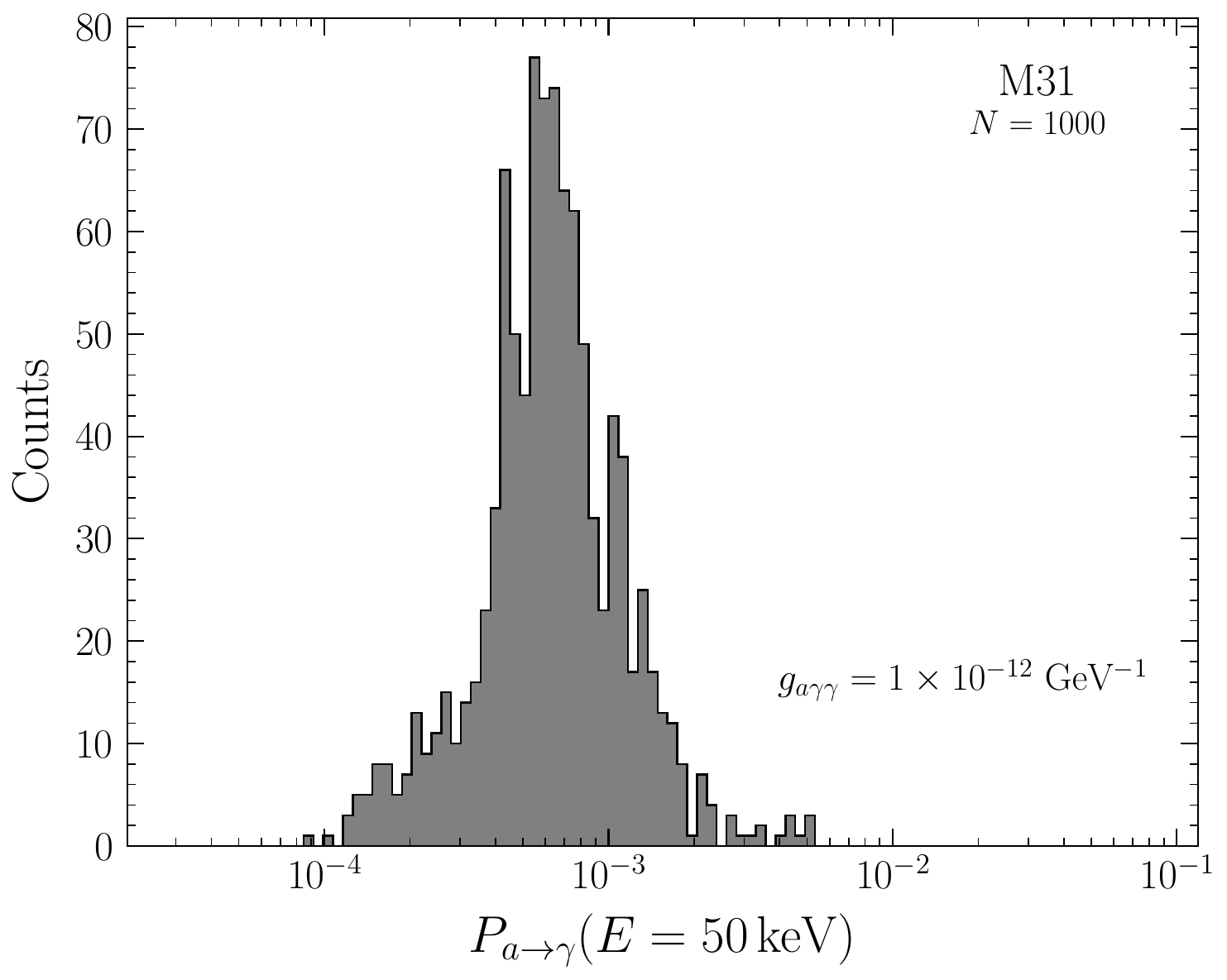}
\caption{Sample distribution of the conversion probabilities for M31 using our fiducial IllustrisTNG TNG50 subhalo and orientation, assuming $g_{a\gamma\gamma} = 10^{-12} \, \, {\rm GeV}^{-1}$ and $m_a \ll 10^{-10}$ eV.  The scatter in this figure arises from the different stellar locations in the galaxy. Results are shown in the low axion mass limit at an energy of 50 keV, which is near the center of our fiducial energy analysis range.}\label{fig:M31_probs}
\end{figure}

In Fig.~\ref{fig:M31_probs}, we show the distribution of conversion probabilities at asymptotically low axion mass and for $E = 50$ keV for M31 for our fiducial candidate and orientation.  The scatter in this distribution arises from the different stellar locations in the galaxy, which are drawn from the baryon distribution.

We finally remark that, in all cases, we additionally include the appropriate Milky Way conversion probabilities~\cite{Unger:2023lob} in our final constraints. This conversion, although highly subdominant in the case of M82 and M87 (as mentioned in~\cite{Ning:2024eky}), has a non-negligible (but still subdominant) effect in M31, where the Milky Way conversion probability is $P_{a\to\gamma} \approx 6.6 \times 10^{-5}$ at $\gagg = 10^{-12}$ GeV$^{-1}$. This can be up to roughly $\sim$10\% of the value of M31's own fiducial $P_{a\to\gamma}$, resulting in an enhancement of the upper limit on $\gaeegagg$ by up to $\sim$4\%.

\section{Comparing Upper Limit Systematics}
\label{app:systematics}

\begin{figure}[!htb]
\centering
\includegraphics[width=\columnwidth]{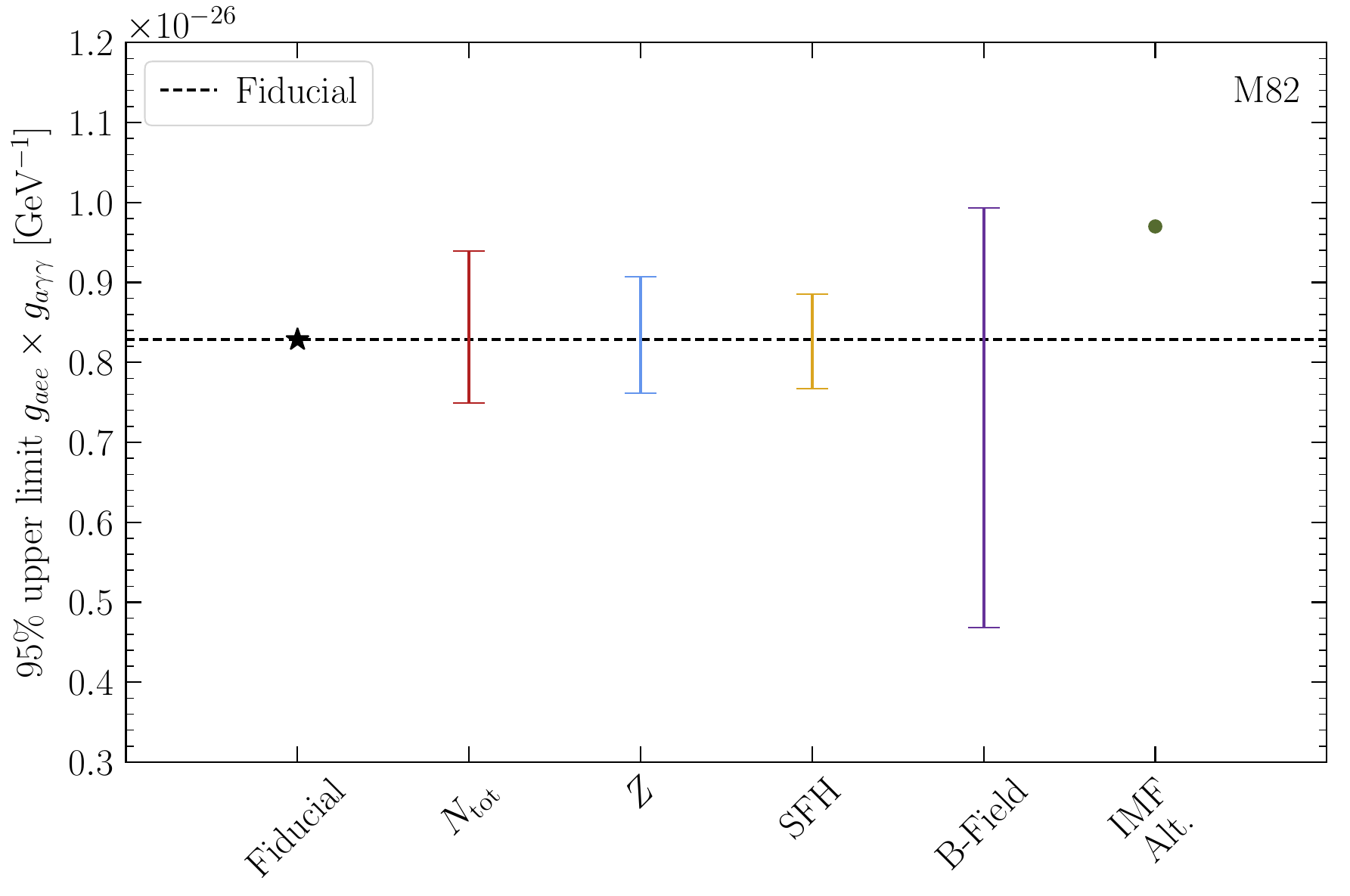}
\caption{The extent of variations in the 95\% upper limits on $\gaeegagg$ in the massless $m_a$ limit, accounting for the major sources of uncertainty in the modeling of M82. We account for the same major uncertainties as in~\cite{Ning:2024eky}: the number of stars ($N_{\rm tot}$), the metallicity ($Z$), the star formation history (SFH), the magnetic field modeling from IllustrisTNG (B-field), and the alternate IMF scenario (IMF Alt.) discussed in~\cite{Ning:2024eky}. The fiducial model in this work returns upper limits given by the black star and the dotted black line. As was true in~\cite{Ning:2024eky}, the dominant source of uncertainty arises from the modeling of magnetic fields in IllustrisTNG.}
\label{fig:M82_systematics}
\end{figure}

\begin{figure}[!htb]
\centering
\includegraphics[width=\columnwidth]{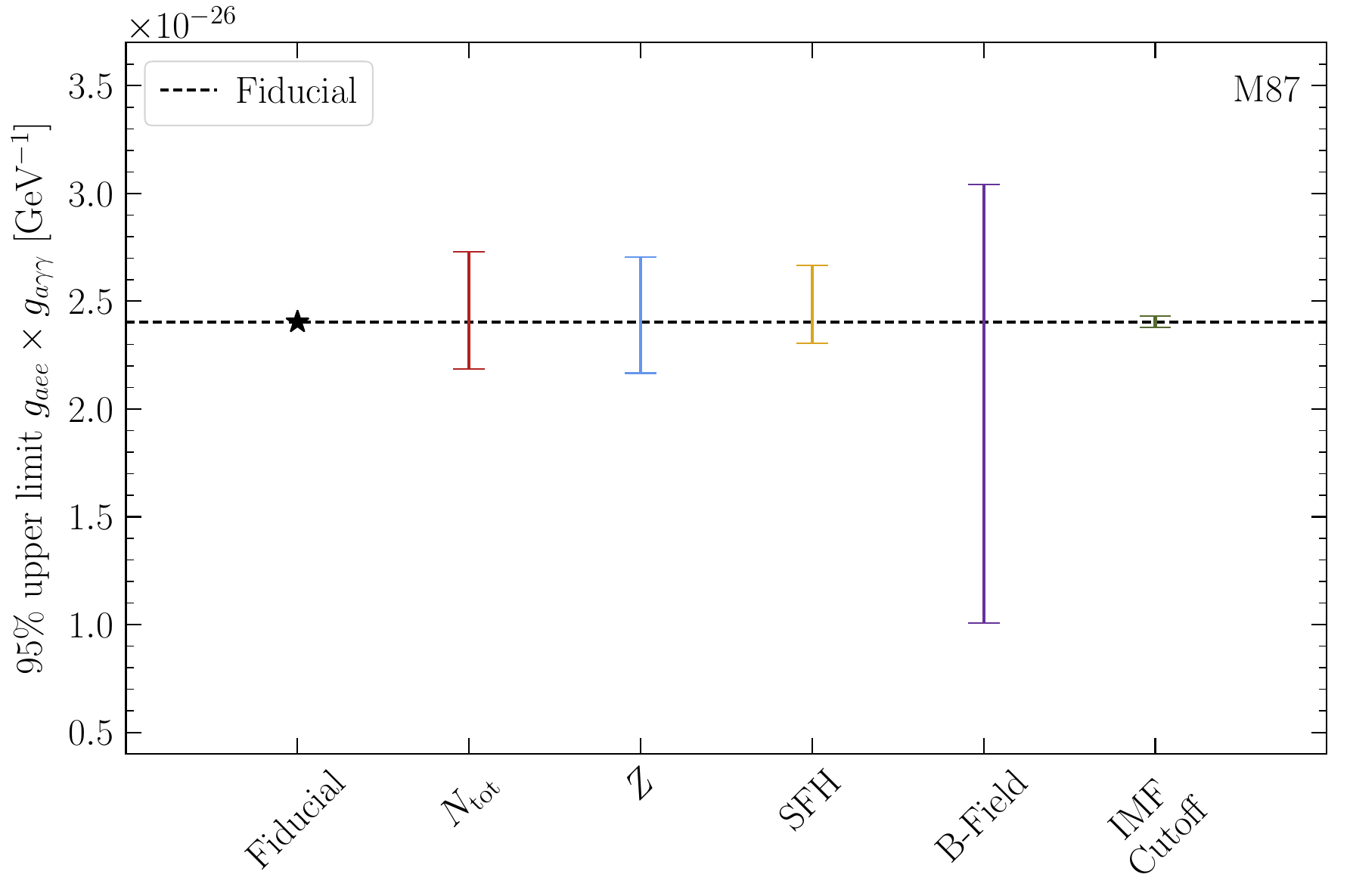}
\caption{The same as Fig.~\ref{fig:M82_systematics} but for M87. Again, we see here that the magnetic field modeling from IllustrisTNG constitutes the dominant source of uncertainty in our analysis.}
\label{fig:M87_systematics}
\end{figure}

\begin{figure}[!htb]
\centering
\includegraphics[width=\columnwidth]{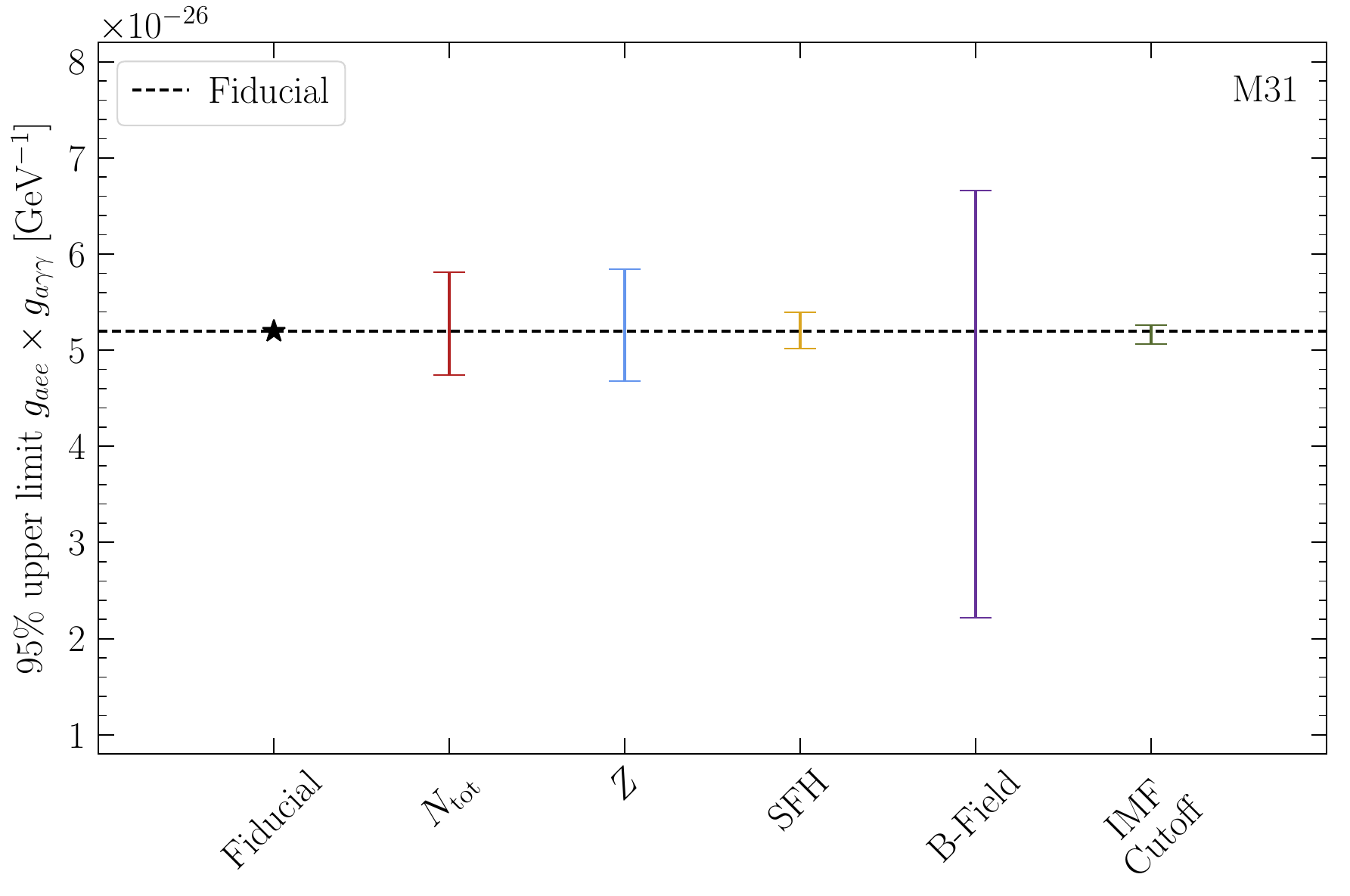}
\caption{The same as Fig.~\ref{fig:M82_systematics} but for M31. Again, the IllustrisTNG magnetic field modeling dominates the uncertainty in our analysis.}
\label{fig:M31_systematics}
\end{figure}

Figs.~\ref{fig:M82_systematics},~\ref{fig:M87_systematics}, and~\ref{fig:M31_systematics} illustrate the approximate extent of the uncertainties in the 95\% upper limits on $\gaeegagg$ as a result of comparing our fiducial model to the major sources of uncertainties in our work for all three of our galaxies M82, M87, and M31. These sources of uncertainty are largely the same as in~\cite{Ning:2024eky}, and include the number of stars, the metallicities, star formation rate, IMF, and magnetic field models sourced from IllustrisTNG. Similar to the results in~\cite{Ning:2024eky}, the dominant sources of uncertainty for all three of the galaxies considered in this work come from the magnetic field models.

\clearpage
\newpage

\bibliography{refs}

\end{document}